\documentclass[twocolumn,showpacs,preprintnumbers,amsmath,amssymb]{revtex4-1}

\usepackage{graphicx}
\usepackage{dcolumn}
\usepackage{bm}
\usepackage{natbib}
\usepackage[usenames,dvipsnames]{xcolor}
\usepackage{pifont}
\usepackage{relsize}
\definecolor{lightgreen}{rgb}{0,1,0}
\definecolor{darkgray}{gray}{0.20}
\usepackage{subfigure}

\begin{document}

\title{Reconstructing the world trade multiplex:\\
the role of intensive and extensive biases}

\author{Rossana Mastrandrea}
\affiliation{Institute of Economics and LEM, Scuola Superiore Sant'Anna, 56127 Pisa (Italy)\\ and Aix Marseille Universit\'e, Universit\'e de Toulon, CNRS, CPT, UMR 7332, 13288 Marseille (France)}
\author{Tiziano Squartini}
\affiliation{Instituut-Lorentz for Theoretical Physics, University of Leiden,  2333 CA Leiden (The Netherlands)\\ and Institute for Complex Systems UOS Sapienza, ``Sapienza''
University of Rome, P.le Aldo Moro 5, 00185 Rome (Italy)}
\author{Giorgio Fagiolo}
\affiliation{Institute of Economics and LEM, Scuola Superiore Sant'Anna, 56127 Pisa (Italy)}
\author{Diego Garlaschelli}
\affiliation{Instituut-Lorentz for Theoretical Physics, University of Leiden,  2333 CA Leiden (The Netherlands)\:}

\date{\today}

\begin{abstract}
In economic and financial networks, the strength 
of each node has always an important economic meaning, such as the size of supply and demand, import and export, or financial exposure. 
Constructing null models of networks matching the observed strengths of all nodes is crucial in order to either detect interesting deviations of an empirical network from economically meaningful benchmarks or reconstruct the most likely structure of an economic network when the latter is unknown.
However, several studies have proved that real economic networks and multiplexes are topologically very different from configurations inferred only from node strengths. 
Here we provide a detailed analysis of the World Trade Multiplex by comparing it to an enhanced null model that 
simultaneously reproduces the strength and the degree of each node.
We study several temporal snapshots and almost one hundred layers (commodity classes) of the multiplex and  find that the observed properties are systematically well reproduced by our model. 
Our formalism allows us to introduce the (static) concept of extensive and intensive bias, defined as a measurable tendency of the network to prefer either the formation of extra links or the reinforcement of link weights, with respect to a reference case where only strengths are enforced.
Our findings complement the existing economic literature on (dynamic) intensive and extensive trade margins. More in general, they show that real-world multiplexes can be strongly shaped by layer-specific local constraints.
\end{abstract}

\maketitle

\section{Introduction}
Over the last fifteen years, there has been a dramatic rise of interest towards the mechanisms of network formation \cite{AlbertBarabasi2002,Jackson03survey}.
One of the reasons 
lies in the fact that the dynamics of a wide range of important phenomena 
(e.g. disease spreading and information diffusion) is strongly affected by the topology of the underlying network mediating the interactions. 
Economic networks are particularly relevant 
for the emergence of many processes of societal relevance such as globalization, economic integration, and financial contagion 
\cite{ScienceNets2009}.

In order to identify the non-trivial structural properties  
of a real network, one needs
to appropriately define and implement a \emph{null model}.
Comparing a real network with 
a null model allows one to 
reveal statistically significant structural patterns.
Substantial effort has been devoted to the 
definition of null models for graphs \cite{Katz_Powell_1957,Erdos_Renyi_1960,Holland_Leinhardt_1976,Rao_etal_1996,Roberts_2000,Newman_2001,Maslov_etal_2004_rewiring_physa,Ansmann_Lehnertz_2011,Bargigli_Gallegati_2011}. 
In economics, the use of purely random models is not new and spans from industrial agglomeration 
\cite{Ellison_Glaeser_1997,Rysman_Greenstein_2005} to international trade  \cite{Armenter_Miklos_2010}. 
Identifying the observed properties of real economic networks 
carrying
non-trivial information 
allows one to select the `target quantities' 
that economic models should aim at explaining or reproducing. 
In particular, if a random network model where only a set of local node-specific properties are enforced turns out to reproduce a real-world network satisfactorily well, then there is no need to know additional information besides the chosen constraints. 
The latter are therefore maximally informative, and the relevant (economic) models should aim at reproducing the observed values of the constraints themselves. On the contrary, a bad agreement indicates the need
for additional information. 

An important economic case study is the World Trade Web (WTW) or International Trade Network (ITN), where nodes are world countries and links represent international trade relationships. 
At the aggregate level, various authors have focused on the binary version of the WTW \cite{SeBo03,Garla2004,Garla2005}, showing (among several other patterns) the presence of a `disassortative' pattern and a stable negative correlation between node degree and clustering coefficient. 
The relevant role played by the topology on the whole network structure is undeniable. In fact, the observed properties turn out to be important in explaining macroeconomics dynamics. For instance, the degree of a country has substantial implications for economic growth and a good potential for explaining episodes of financial contagion \cite{KaliReyes2007, Kali_Reyes_2010_contagion}. 

The introduction of null models in the analysis of the WTW  
has allowed to (re)interpret the results of the network approach in the light of traditional economic analyses \cite{Feenstra2004}, where most country-specific macroeconomic variables 
(total trade, number of trade partners, etc.) can be rephrased in terms of first-order (local) properties of nodes. 
Recent studies \cite{Squartini_etal_2011a_pre,fagiolo2013null,squartini2012triadic} have shown that much of the binary architecture of the WTW can be reproduced by a null model controlling for the (in- and out-) degrees of all countries. 
This null model is known as the Binary Configuration Model (BCM).
The agreement between the BCM and the real WTW holds both at the aggregate level and at a `multiplex' \cite{multiplex_prl,multiplex_ginestra,multiplex_porter,multiplex_latora} level (i.e. for every different commodity-specific layer of the network) \cite{Squartini_etal_2011a_pre,fagiolo2013null}.
This result has important consequences for trade modelling, because it highlights the need for revised macroeconomic theories aimed at reproducing the degrees of all countries, in contrast with the standard approach which considers the number of trade partners irrelevant.

Despite its fundamental role, the binary version of the WTW suffers from an important limitation: it ignores the magnitude of the observed trade relationships, thus giving only partial information about the network. 
Indeed, it turns out that the binary and weighted versions of the WTW are very different 
\cite{Fagiolo2008physa,Fagiolo2009pre,Fagiolo2010jeic}. For example, the edge weight and node strength distributions are highly left-skewed, indicating that few intense trade connections co-exist with a majority of low-intensity ones \cite{Bhatta2007a} and that countries with many partners are also the richest and most (globally) central. However, the latter trade very intensively with only few of their partners (which turn out to be themselves very connected), forming few but intensive trade clusters (i.e. triangular patterns) \cite{Fagiolo2009pre}. 

Recent works \cite{Squartini_etal_2011b_pre,fagiolo2013null} have extended the binary null model analysis \cite{Squartini_etal_2011a_pre} to the weighted version of the WTW, now randomizing the latter while preserving each country's (in- and out-)strength.
This null model is known as the Weighted Configuration Model (WCM). Surprisingly, these studies found a very bad agreement between observed and expected higher-order weighted 
properties, mainly because the randomized network is in general much denser than the observed one, and often  almost fully connected. 
Other works obtained similar results using different methodologies  \cite{serrano2005weighted,serrano2006correlations,Ansmann_Lehnertz_2011,fronczak2012statistical,Bhatta2007a}.

The standard interpretation of these findings calls for the existence of higher-order mechanisms shaping the structure of the WTW as a weighted network, suggesting the impossibility to reconstruct the WTW from purely country-specific information.
Again, this interpretation has important consequences for economic modelling. In fact, reproducing the observed strengths or other purely weighted properties which, unlike the degrees, represent the main target of traditional macroeconomic theories (the most popular example being that of Gravity Models \cite{Bergstrand1985,chaney2008distorted,duenas2011modeling,myTinbergen}) appears quite useless in order to explain the network structure.
Even if for the opposite reason, this conclusion calls again for a change of perspective in the way economic models approach the international trade system.
However, in this paper we show that this interpretation should be considerably revised. 

There is another attractive reason for using null models in network analysis, namely the possibility of reconstructing an (unobserved) network from its local properties. 
We recently introduced an enhanced method to build ensembles of networks simultaneously reproducing the strength and the degree of each node \cite{mastrandrea2013enhanced}. 
The method, called Enhanced Configuration Model (ECM) because it is an augmented version of both the BCM and the WCM, builds upon prior theoretical results introducing the generalized Bose-Fermi distribution \cite{mybosefermi}, which is the appropriate probability function characterizing systems with both binary and non-binary constraints.
The application of the ECM allowed us to show that, for many real networks where the specification of the strengths alone give very poor results, the joint specification of strengths and degrees can reconstruct the original network to a great degree of accuracy \cite{mastrandrea2013enhanced}.

While we already analysed one (aggregated and static) snapshot of the WTW as part of the analysis described above, this is not enough in order to conclude that those results can be straightforwardly extended to different temporal snapshots and different layers of this economic multiplex.
Indeed, ref. \cite{mastrandrea2013enhanced} focused on a diverse set of networks of very different nature (biological, social, etc.), but included no temporal analysis and no multiplex analysis. 
As we have already mentioned, in the particular case of the WTW carrying out a deeper analysis has a special importance for its macroeconomic implications, because understanding the empirical role of local constraints changes the theoretical approaches to the trade system and can highlight some important flaws in standard macroeconomic modeling. 
Previous investigations of the international trade network revealed similarities but also differences across the layers of the multiplex (at both binary and weighted levels) \cite{BariFagiGarla2010}.
Moreover, increased product complexity appears to yield increased product-specific network complexity \cite{debene_sectoral}.
Finally, the country-product associations reveal a highly nontrivial nested structure \cite{pietronero} which contradicts the mainstream economic expectation and, similarly, the product-product network displays a strong core-periphery structure \cite{hidalgo}.
These product-related heterogeneities imply that it is not obvious whether one should expect that the results obtained on a single, aggregate instance of the WTW are robust over time and across different commodity classes. 

Given the importance of the problem and its more general implications for the understanding of economic networks, in this paper we carry out an in-depth investigation of the WTW spanning several years and different commodities.
Our results show that, when considered together, the total trade (strength) and the number of trade partners (degree) of all world countries are enough in order to reproduce many higher-order properties of the network, for all levels of disaggregation and all temporal snapshots in our analysis. In order to fully explain the structure of the WTW, binary constraints must therefore be added to the weighted ones. Economically, this suggests that additional higher-order mechanisms besides those accounting for the joint evolution of degrees and strengths are not really necessary in order to explain the structure and dynamics of the WTW. 
On the other hand, we also show that degrees and strengths are `irreducible' to each other, i.e. any minimal macroeconomic model aiming at reproducing the WTW should not discard any of the two quantities.

Our work complements the existing economic literature about the so-called \emph{extensive} and \emph{intensive margins} of trade \cite{ricardo1819principles, felbermayr2006exploring,DeBene_Tajoli_2011}, defined as the preference for the network to evolve by establishing new connections or by strengthening the intensity of existing ones respectively. 
While extensive and intensive margins are traditionally defined at an intrinsically dynamical level, we define the new concept of \emph{extensive and intensive biases} as a purely static notion, and for each pair of countries separately. We focus on cross-sectional data and evaluate whether, at a given point in time, pairs of countries are `shifted' along the intensive or extensive direction as compared to an appropriate null model. 
Thus, our methodology does \emph{not} 
establish whether the WTW proceeds along the extensive or intensive margin in the traditional `dynamical' way, i.e. by accounting for the variation of trade connections and/or their weights over time. 
On the contrary, it allows us to identify a bias towards either the extensive or the intensive direction in a novel, static fashion, by exploiting a mathematical property of the null model specifying both strengths and degrees. Moreover, the entity of the bias can be exactly quantified for each pair of countries, allowing different pairs of nodes to be characterized by opposite tendencies. 

Our paper is organized as follows. In Sec. \ref{sec:data} we introduce the trade data used and briefly summarize the methodology we employ to specify both strengths and degrees in weighted networks \cite{Squartini_Garlaschelli_2011,mastrandrea2013enhanced}. In Sec. \ref{result} we apply the methodology first to a reference year, then to several temporal snapshots, and finally to different aggregation levels (commodity classes) of the world trade multiplex. 
In Sec. \ref{sec:discussion} we discuss some general economic implications of our results and their relation with the more traditional literature about intensive and extensive margins of trade.


\section{Data and Methodology\label{sec:data}}

As we mentioned, our analysis builds upon previous studies of the WTW that showed that the degree sequence is able to replicate the purely binary topology very well \cite{Squartini_etal_2011a_pre}, while the strength sequence is completely unable to replicate the weighted structure \cite{Squartini_etal_2011b_pre}.
We aim at understanding whether the joint specification of strengths and degrees allows us to successfully replicate the weighted structure. 
To this end, we use exactly the same data set as refs. \cite{Squartini_etal_2011a_pre,Squartini_etal_2011b_pre}, so that consistent comparisons can be made. 
Such data set is described in sec. \ref{sec:data1}.
We also use a similar, but appropriately generalized, methodology.
This is introduced in sec. \ref{sec:data2}.

\begin{table*}
\center \small
\begin{tabular}{cp{6cm}ccc}
\hline
\hline
   HS Code &  Commodity & Value (USD) & Value per link (USD) & $\%$ of aggregate trade \\
\hline
        84 & Nuclear reactors, boilers, machinery and mechanical appliances; parts thereof  &   $5.67\times 10^{11}$ &   $6.17\times 10^{7}$ &    11.37\% \\

        85 & Electric machinery, equipment and parts; sound equipment; television equipment  &   $5.58\times 10^{11}$ &   $6.37\times 10^{7}$ &    11.18\% \\

        27 & Mineral fuels, mineral oils \& products of their distillation; bitumin substances; mineral wax  &   $4.45\times 10^{11}$ &   $9.91\times 10^{7}$ &     8.92\% \\

        87 & Vehicles, (not railway, tramway, rolling stock); parts and accessories  &   $3.09\times 10^{11}$ &   $4.76\times 10^{7}$ &     6.19\% \\

        90 & Optical, photographic, cinematographic, measuring, checking, precision, medical or surgical instruments/apparatus; parts \& accessories  &   $1.78\times 10^{11}$ &   $2.48\times 10^{7}$ &     3.58\% \\

        39 & Plastics and articles thereof.  &   $1.71\times 10^{11}$ &   $2.33\times 10^{7}$ &     3.44\% \\

        29 & Organic chemicals  &   $1.67\times 10^{11}$ &   $3.29\times 10^{7}$ &     3.35\% \\

        30 & Pharmaceutical products  &    $1.4\times 10^{11}$ &   $2.59\times 10^{7}$ &     2.81\% \\

        72 & Iron and steel  &   $1.35\times 10^{11}$ &   $2.77\times 10^{7}$ &     2.70\% \\

        71 & Pearls, precious stones, metals, coins, etc &   $1.01\times 10^{11}$ &   $2.41\times 10^{7}$ &     2.02\% \\

        10 &   Cereals  &   $3.63\times 10^{10}$ &   $1.28\times 10^{7}$ &     0.73\% \\

        52 & Cotton, including yarn and woven fabric thereof  &   $3.29\times 10^{10}$ &   $6.96\times 10^{6}$ &     0.66\% \\

         9 & Coffee, tea, mate \& spices  &   $1.28\times 10^{10}$ &   $2.56\times 10^{6}$ &     0.26\% \\

        93 & Arms and ammunition, parts and accessories thereof &   $4.31\times 10^{9}$ &   $2.46\times 10^{6}$ &     0.09\% \\

       ALL &  Aggregate (all 97 commodities) &   $4.99\times 10^{12}$ &   $3.54\times 10^{8}$ &   100.00\% \\
       \hline
\hline
\end{tabular}
\caption{The $14$ most relevant commodity classes (plus aggregate trade) in year 2003 and the corresponding total trade value (USD), trade value per link (USD), and share of world aggregate trade.
From ref. \cite{BariFagiGarla2010}.\label{tab}}
\end{table*}

\subsection{The world trade multiplex \label{sec:data1}}

As in refs. \cite{Squartini_etal_2011a_pre,Squartini_etal_2011b_pre}, we employ international trade data provided by the United Nations Commodity Trade Database (UN COMTRADE \footnote{http://comtrade.un.org/}) in order to build a time sequence of binary and weighted trade multiplexes. The sample refers to $11$ years (1992-2002) and the money unit is current U.S. dollars. 
The choice of this time span allows us construct a time-varying multiplex with a constant number of $N=162$ countries and a constant number of $C=97$ layers (commodity classes), evolving over $T=11$ years. 
When necessary, we will also aggregate all or part of the layers to obtain different levels of resolution.

We chose the classification of trade values into $C=97$ possible commodities listed according to the Harmonized System 1996 (HS1996 \footnote{http://unstats.un.org/}). For each year $t$ ($1\le t\le T$) and each commodity $c$ ($1\le c\le C$), the starting data are represented as a matrix whose elements are the trade flows directed from each country to all other countries \footnote{We employ the flow as reported by the importer \cite{Fagiolo2008physa} since the importer and exporter records do not always match.}.
The matrix elements specifying the multiplex are denoted by $e^{(c)}_{ij}(t)$, where $e^{(c)}_{ij}(t)> 0$ whenever there is an export of good $c$ from country $i$ to country $j$, and $e^{(c)}_{ij}(t)= 0$ otherwise.
Rows and columns stand for  exporting and importing countries respectively.
The value of $e^{(c)}_{ij}(t)$ is in current U.S. dollars (USD) for all commodities. 

Given the commodity-specific (multiplex) data $e_{ij}^{(c)}(t)$, we can compute the total (aggregate) value of exports $e_{ij}^{AGG}(t)$ from country $i$ to country $j$ summing up over the exports of all $C=97$ commodity classes:
\begin{equation}
e_{ij}^{AGG}(t)\equiv \sum_{c=1}^C e_{ij}^{(c)}(t)
\end{equation}
The particular aggregation procedure described above, introduced in \cite{BariFagiGarla2010}, allows us to  compare the analysis of the $C$ commodity-specific layers of the multiplex with a {$(C+1)$-th} aggregate network, avoiding possible inconsistencies between the original aggregated and disaggregated trade data. 

We put special emphasis on the $14$ particularly relevant commodities identified in \cite{BariFagiGarla2010} and reported in table \ref{tab}. They include the $10$ most traded commodities ($c = 84, 85, 27, 87, 90, 39, 29, 30, 72, 71$ according to the HS1996) in terms of total trade value (following the ranking in year 2003, \cite{BariFagiGarla2010}), plus $4$ commodities ($c = 10, 52, 9, 93$ according to the HS1996) which are less traded but still important for their economic relevance. 
The 10 most traded commodities account for $56\%$ of total world trade in 2003; moreover, they also feature the largest values of trade value per link (see table \ref{tab}). Taken together, the $14$ most relevant commodities account in total for $57\%$ of world trade in 2003. As an intermediate level of aggregation between individual commodities and fully aggregate trade, we also consider the network formed by the sum of the 14 special commodities. 
In this way we can also draw conclusions about the robustness of our methodology with respect to  different levels of aggregation. 

In our analyses, we will focus on the undirected (symmetrized) representation of the network for obvious reasons of simplicity, even if the extension to the directed case is straightforward once the method in ref. \cite{mastrandrea2013enhanced} is appropriately generalized. 
In any case, several works have shown that the percentage of reciprocated interactions in the WTW is steadily high \cite{Squartini_etal_2011a_pre,Squartini_etal_2011b_pre,Fagiolo2008physa}, giving us reasonable confidence that we can focus on the temporal series of undirected networks.
We therefore define the symmetric matrices
\begin{eqnarray}
\tilde{w}_{ij}^{(c)}(t) &\equiv& \left\lfloor\frac{e_{ij}^{(c)}(t)+e_{ji}^{(c)}(t)}{2}\right\rceil,\notag\\ 
\tilde{w}_{ij}^{AGG}(t) &\equiv& \left\lfloor\frac{e_{ij}^{AGG}(t)+e_{ji}^{AGG}(t)}{2}\right\rceil 
\end{eqnarray}
where $\lfloor x \rceil$ represents the nearest integer to the nonnegative real number $x$ \footnote{Rounding to integers is required by the randomization procedure (for more details see \cite{Squartini_Garlaschelli_2011} and \cite{mastrandrea2013enhanced}).}.
The above matrices define an undirected weighted network where the weight of a link is the average of the trade flowing in either direction between two countries.

In order to wash away trend effects and make the data comparable over time, we normalized our weights according to the total trade volume for each year:
\begin{equation}
w_{ij}^{(c)}(t)\equiv \frac{\tilde{w}_{ij}^{(c)}(t)}{\tilde{w}_{TOT}^{(c)}}, \qquad w_{ij}^{AGG}(t)\equiv \frac{\tilde{w}_{ij}^{AGG}(t)}{\tilde{w}_{TOT}^{AGG}}
\end{equation}
where $\tilde{w}_{TOT}^{(c)}=\sum_{i=1}^N \sum_{j=i+1}^N \tilde{w}_{ij}^{(c)}$ and $\tilde{w}_{TOT}^{AGG}=\sum_{i=1}^N \sum_{j=i+1}^N \tilde{w}_{ij}^{AGG}$. 
In such a way, we end up with adimensional weights that allow proper comparisons over time and consistent analyses of the evolution of network properties. 

\subsection{The Enhanced Configuration Model\label{sec:data2}}

Our methodology makes intense use of the ECM \cite{mastrandrea2013enhanced}, defined as an ensemble of weighted networks with given strengths and degrees. 
In some sense, the ECM unifies the BCM and the WCM, which have been separately used in previous analyses of the same data \cite{Squartini_etal_2011a_pre,Squartini_etal_2011b_pre}.

One can show \cite{Squartini_Garlaschelli_2011,mymaxsam} that most of the algorithms developed to randomize a real-world network suffer from severe limitations and give biased results. 
To overcome these limitations, we use a recently proposed unbiased method \cite{mastrandrea2013enhanced} based on the maximum-likelihood estimation \cite{mylikelihood} of maximum-entropy ensembles of graphs \cite{Squartini_Garlaschelli_2011}. 
One of the attractive features of this method is its fully analytical character, which allows us to obtain the exact expressions for the expected values without performing explicit averages over numerically sampled networks of the ensemble.
Recently, a fast and unbiased algorithm has been released to computationally implement this procedure \cite{mymaxsam}.
We briefly recall the main steps of this approach and of the enhanced network reconstruction method that, building on some general theoretical results \cite{mybosefermi}, can be derived from it.

Firstly, we need to specify a set of constraints $\{C_i\}$. 
The constraints are the network properties that we want to preserve during the randomization procedure, according to the specific network and research question. Generally these constraints are \emph{local}, such as the strength sequence defining the WCM, but the methodology can also account for non-local constraints in some cases \cite{squartini2012reciprocity,picciolo2012role}. 
In order to construct the ECM, defined an ensemble of weighted networks where both the degree sequence and the strength sequence are specified \cite{mastrandrea2013enhanced}, we choose  
\begin{equation}\label{cons}
\{C_i\}\equiv\{k_i,s_i\}
\end{equation}
where $k_i$ stands for the $i$-th node degree and $s_i$ for the $i$-th node strength.

Secondly, we need to find the analytical expression for the probability $P(W)$ that (under the chosen constraints) maximizes the Shannon-Gibbs entropy
\begin{equation}
S(W)\equiv -\sum_W P(W) \ln P(W)
\end{equation}
over the ensemble of allowed networks.
Note that $P(W)$ stands for the occurrence probability of the graph $W$ in the ensemble of allowed weighted graphs, and the sums are over all such graphs. 
For our purposes, the allowed graphs are all the undirected networks with $N$ vertices and non-negative integer edge weights.
Each such network is uniquely specified by its $N\times N$ symmetric weight matrix $W$, where the entry $w_{ij}=w_{ji}\in\mathbb{N}$ represents the weight of the link connecting nodes $i$ and $j$.
The maximization of Shannon's entropy is done under the constraints $\sum_W P(W) =1$ (this ensures the normalization of the probability) and $\langle C_i\rangle\equiv\sum_W P(W)C_i(W)=C_i$ for all $i$ (this fixes the desired structural properties).
The formal solution \cite{Squartini_Garlaschelli_2011} of this constrained maximization problem can be written as 
\begin{equation}
P(W|\vec{\theta})\equiv \frac{e^{-H(W,\vec{\theta})}}{Z(\vec{\theta})},
\end{equation}
where 
\begin{equation}
H(W,\vec{\theta}) \equiv \sum_i \theta_i C_i(W)
\end{equation} 
is the \emph{graph Hamiltonian} and 
\begin{equation}
Z(\vec{\theta})\equiv \sum_W e^{-H(W,\vec{\theta})}
\end{equation}
is the \emph{partition function}.
The Hamiltonian is a linear combination of the constraints $\{C_i\}$, with the coefficients $\{\theta_i\}$  being the conjugate Lagrange multipliers introduced in the constrained-maximization problem.

For the ECM, it is possible to show \cite{mastrandrea2013enhanced,mybosefermi} that
\begin{equation}\label{maxlik}
P(W|\vec{x},\vec{y})=\prod_{i<j}q_{ij}(w_{ij}|\vec{x},\vec{y}),
\end{equation}
where $\vec{x}$ and $\vec{y}$ are two $N$-dimensional Lagrange multipliers ($N$ stands for the number of nodes) controlling for the expected degrees and strengths respectively (with $x_i\ge 0$ and $0\le y_i<1$ for all $i$) and $q_{ij}(w|\vec{x},\vec{y})$ is the conditional probability to observe a link of weight $w$ between nodes $i$ and $j$. The latter has the explicit expression \cite{mybosefermi,mastrandrea2013enhanced}
\begin{equation}\label{qw}
q_{ij}(w|\vec{x},\vec{y})=\frac{(x_{i}x_{j})^{\Theta(w)}(y_{i}y_{j})^{w}(1-y_{i}y_{j})}{1-y_{i}y_{j}+x_{i}x_{j}y_{i}y_{j}}.
\end{equation}

The third step of the procedure prescribes to find the values of the Lagrange multipliers $\vec{x}^*,\vec {y}^*$ that maximize the log-likelihood of generating the observed weighted network $W^*$ (which is one particular network in the ensemble considered).
The log-likelihood reads
\begin{equation}
\mathcal{L}(\vec{x},\vec{y})\equiv
 \ln P(W^*|\vec{x},\vec{y})=\sum_{i<j}\ln q_{ij}(w^*_{ij}|\vec{x},\vec{y})
\end{equation}
representing the logarithm of the probability to observe the empirical graph $W^*$. The maximization of the likelihood is equivalent to the requirement that the desired constraints are satisfied on average by the ensemble of networks \cite{mylikelihood}, i.e. in this case $\langle k_i\rangle = k_i(W^*)$ and $\langle s_i\rangle = s_i(W^*)$ for all $i$ \cite{mastrandrea2013enhanced}.

As a final step, one can use the Lagrange multipliers  $\vec{x}^*,\vec {y}^*$ to compute the expected value $\langle X \rangle$ of any (higher-order) network property $X(W)$:
\begin{equation}\label{avpr}
\langle X \rangle \equiv \sum_W X(W)P(W|\vec{x}^*,\vec{y}^*)
\end{equation}
Comparing $\langle X \rangle$ with the observed value $X(W^*)$ allows us to verify whether the `reconstructed' value of the property is indeed close to the empirical one.

We will also compare the predictions of the ECM with those of the WCM. The latter can be obtained by setting $\vec{x}=\vec{1}$ and maximizing the likelihood with respect to $\vec{y}$ alone, thus finding another vector $\vec{y}^{**}\ne\vec{y}^{*}$ \cite{mastrandrea2013enhanced}.

A computationally fast and statistically unbiased algorithm to obtain the values of the Lagrange multipliers maximizing the likelihood of both the ECM and WCM (along with other maximum-entropy ensembles) has been recently introduced under the name of `Max \& Sam' (`maximize and sample') method \cite{mymaxsam}.
We will use that algorithm in our analysis.


\section{Results}\label{result}

In this section, we first analyse in detail the aggregated trade network in the reference year 2002. 
Our choice of this particular snapshot is dictated by the need to compare our results with that of refs. \cite{Squartini_etal_2011b_pre,Squartini_etal_2011a_pre}, as we mentioned.
We then consider the temporal evolution of the system. Finally, we perform a multiplex analysis on the disaggregated, commodity-specific layers of the WTW. 

\subsection{Analysis of the aggregated network}\label{wtwagg}

For simplicity of the notation, in this subsection we indicate with $A$ the adjacency matrix and with $W$ the weighted matrix representing the aggregate network in year 2002, i.e. $w_{ij}\equiv w_{ij}^{AGG}(2002)$.
In general, we can relate the entries of the matrices $A$ and $W$ through the zeroth power using the notation $w^0_{ij}=a_{ij}$, where we conventionally define $0^0\equiv0$. 
This notation will be useful later to calculate the expected values of various structural properties.
It also allows us to express the  properties of purely topological properties, which in principle depend on the adjacency matrix $A$, as functions of the matrix $W$.

We are interested in assessing to what extent the ECM is able to replicate the higher-order properties characterizing the WTW over time. 
Therefore we first apply the ECM to the data, thus finding the vectors $\vec{x}^*$ and $\vec{y}^*$, and then we use these vectors to calculate the expected values of the chosen higher-order properties.
For consistency with refs. \cite{Squartini_etal_2011a_pre,Squartini_etal_2011b_pre}, we focus on the Average Nearest Neighbor Degree, the Average Nearest Neighbor Strength (denoted by $k^{nn}_i$ and $s^{nn}_i$ respectively), the Binary Clustering Coefficient and the Weighted Clustering Coefficient (denoted by $c_i$ and $c^w_i$ respectively). 
In what follows, we first recall the analytical expressions of these higher-order quantities. 
Then, for the sake of clarity we write down the explicit expressions for the expected value of the same quantities under the ECM, i.e. the particular form taken by eq.\eqref{avpr} for each property under study.

The Average Nearest Neighbor Degree is defined as
\begin{equation}\label{ANND}
k^{nn}_i(W)\equiv\frac{\sum_{j\ne i}w^0_{ij}k_j}{k_i}=\frac{\sum_{j\ne i}\sum_{k\ne j}w^0_{ij}w^0_{jk}}{\sum_{j\ne i} w^0_{ij}},
\end{equation}
where $k_i=\sum_{j\ne i} w^0_{ij}$ stands for the $i$-th node degree,
and represents the average of the degrees of the partners of node $i$, i.e. it provides a measure of the connectivity of the neighbours of that node.

The Binary Clustering Coefficient has the expression
\begin{equation}\label{BCC}
c_i(W) \equiv\frac{\sum_{j\neq i}\sum_{k\ne i,j}w^0_{ij}w^0_{jk}w^0_{ki}}{\sum_{j\neq i}\sum_{k\ne i,j}w^0_{ij}w^0_{ki}}
\end{equation}
and measures the tendency of node $i$ to form triangles, i.e. it counts how many closed triangles are attached to node $i$, divided by the maximum number of triangles achievable by a node with degree $k_i$. 

The Average Nearest Neighbor Strength is defined as
\begin{equation}
{s}^{nn}_i(W)\equiv\frac{\sum_{j\ne i}w^0_{ij}{s}_j}{k_i}=\frac{\sum_{j\ne i}\sum_{k\ne j}w^0_{ij}w_{jk}}{\sum_{j\ne i} w^0_{ij}},
\label{ANNS}
\end{equation}
where $s_i=\sum_{j\ne i} w_{ij}$ stands for the $i$-th node strength,
and measures the average strength of the neighbors of vertex $i$. Similarly to its binary counterpart ($k_i^{nn}$), ${s}^{nn}_i$ reveals the `intensity' of connectiviy of the neighbours of a node, now taking edge weights into account. 

Finally, the Weighted Clustering Coefficient \cite{Fagiolo2007pre} can be defined as
 \begin{equation}
{c}_i^w(W)=\frac{\sum_{j\neq i}\sum_{k\ne i,j}({w}_{ij}{w}_{jk}{w}_{ki})^{1/3}}{\sum_{j\neq i}\sum_{k\ne i,j}w^0_{ij}w^0_{ki}}
\label{WCC}
\end{equation}
and measures the propensity of node $i$ to be involved in triangular relations, taking into account the weights of such relations. 

In order to compute the expected value of the above properties, it is necessary to compute the expected product of (powers of) distinct matrix entries.
The independence of pairs of nodes in the ECM \cite{mastrandrea2013enhanced} ensures that 
\begin{equation}
\left\langle \sum_{i\ne j\ne k,\dots}w_{ij}^{\alpha}\cdot w_{jk}^{\beta}\cdot\:\dots\right\rangle=\sum_{i\ne j\ne k,\dots}\langle w_{ij}^{\alpha}\rangle \cdot\langle w_{jk}^{\beta}\rangle\cdot\:\langle\dots\rangle.
\nonumber
\end{equation}
Each individual term in the product is given by
\begin{equation}
\langle w_{ij}^{\gamma}\rangle=\sum_{w=0}^{+\infty}\! w^{\gamma}q_{ij}(w|\vec{x}^*,\vec{y}^*)=\frac{x^*_{i}x^*_{j}(1-y^*_{i}y^*_{j})\mbox{Li}_{-\gamma}(y^*_{i}y^*_{j})}{1-y^*_{i}y^*_{j}+x^*_{i}x^*_{j}y^*_{i}y^*_{j}}
\nonumber
\end{equation}
where $\mbox{Li}_{n}(z)\equiv\sum_{l=1}^{+\infty}z^l/l^n$ is the $n$th polylogarithm of $z$ \cite{mastrandrea2013enhanced}. The simplest cases $\gamma =1$ and $\gamma=0$ yield the expected weight 
\begin{equation}
\langle w_{ij} \rangle\equiv \frac{x^*_i x^*_j y^*_i y^*_j }{(1-y^*_i y^*_j)(1-y^*_i y^*_j+x^*_i x^*_jy^*_iy^*_j)}
\end{equation}
and the connection probability 
\begin{equation}
p_{ij}\equiv \langle w^0_{ij}\rangle=\frac{x^*_i x^*_j y^*_i y^*_j }{1-y^*_i y^*_j+x^*_i x^*_jy^*_iy^*_j}
\label{pij}
\end{equation}
respectively \cite{mastrandrea2013enhanced}.

As a result, the expected values of the purely topological (weight-independent) properties can be obtained by simply replacing $w^0_{ij}=a_{ij}$ with $p_{ij}$  \cite{mastrandrea2013enhanced}: 
\begin{equation}\label{ANNDe}
\langle k^{nn}_i(W) \rangle \equiv\frac{\sum_{j\ne i}p_{ij} \langle k_j \rangle}{\langle k_i\rangle}=\frac{\sum_{j\ne i}\sum_{k\ne j}p_{ij}p_{jk}}{\sum_{j\ne i} p_{ij}},
\end{equation}
\begin{equation}\label{BCCe}
\langle c_i(W) \rangle \equiv\frac{\sum_{j\neq i}\sum_{k\ne i,j}p_{ij}p_{jk}p_{ki}}{\sum_{j\neq i}\sum_{k\ne i,j}p_{ij}p_{ki}}
\end{equation}
where, by construction, $\langle k_i\rangle \equiv k_i$ for all $i$. 
For the expected value of ${s}^{nn}_i$ we have \cite{mastrandrea2013enhanced,mybosefermi}: 
\begin{equation}
\langle {s}^{nn}_i(W) \rangle \equiv\frac{\sum_{j\ne i}p_{ij}{\langle s}_j\rangle}{\langle k_i\rangle}=\frac{\sum_{j\ne i}\sum_{k\ne j}p_{ij}\langle w_{jk}\rangle}{\sum_{j\ne i} p_{ij}}
\end{equation}
where $\langle s_i \rangle \equiv s_i$, $\forall i$.
Finally, the expected value of $c^w_i$ is
\begin{equation}
\langle{c}_i^w(W)\rangle=\frac{\sum_{j\neq i}\sum_{k\ne i,j}
\langle{w}_{ij}^{1/3}\rangle
\langle{w}_{jk}^{1/3}\rangle
\langle{w}_{ki}^{1/3}\rangle}{\sum_{j\neq i}\sum_{k\ne i,j}p_{ij}p_{ki}}
\label{eq:cw}
\end{equation}

\begin{figure*}[!t]
\centering
{\includegraphics[width=0.4\textwidth]{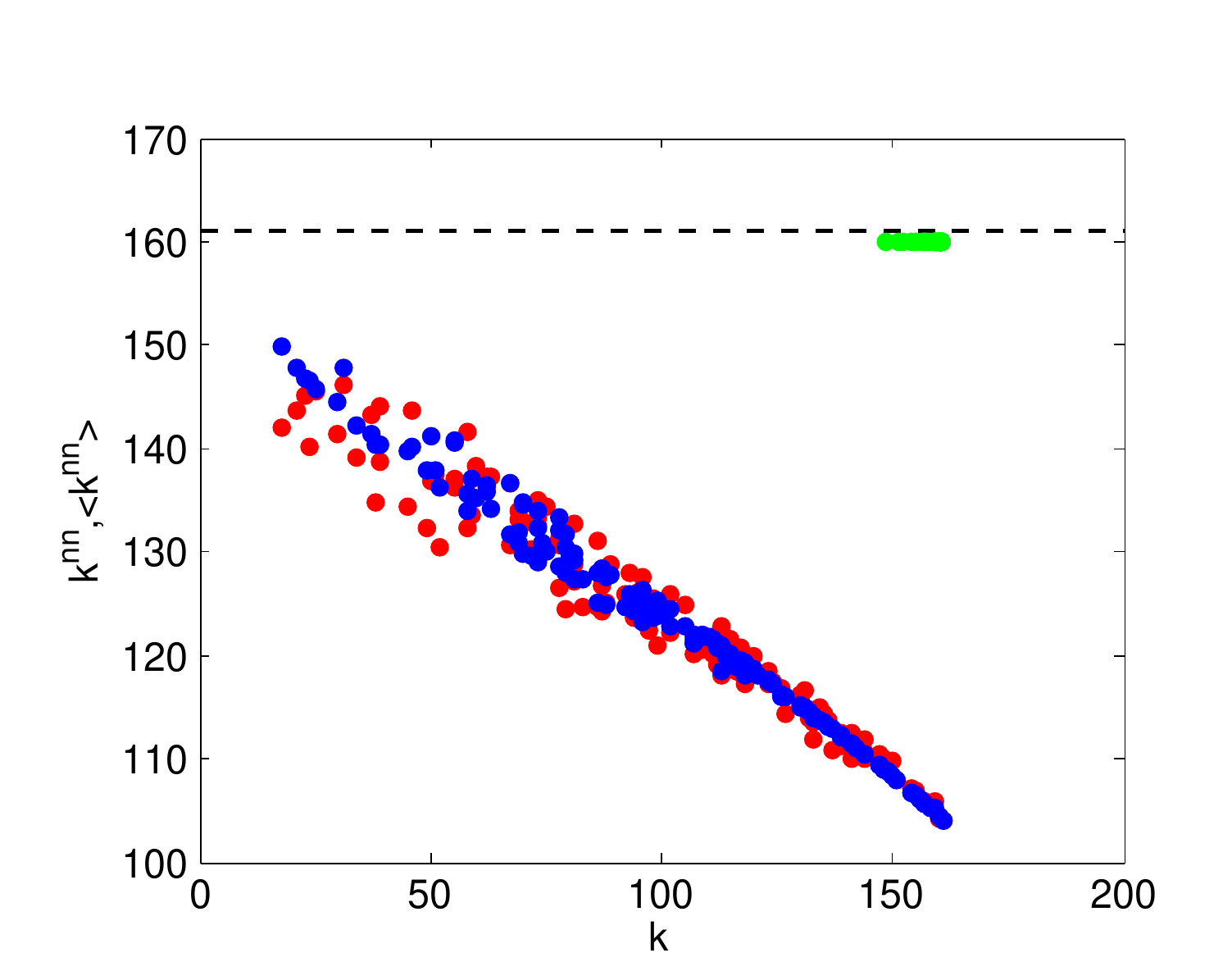}}
\hspace{-5mm}
\vspace{-10mm}
 {\includegraphics[width=0.4\textwidth]{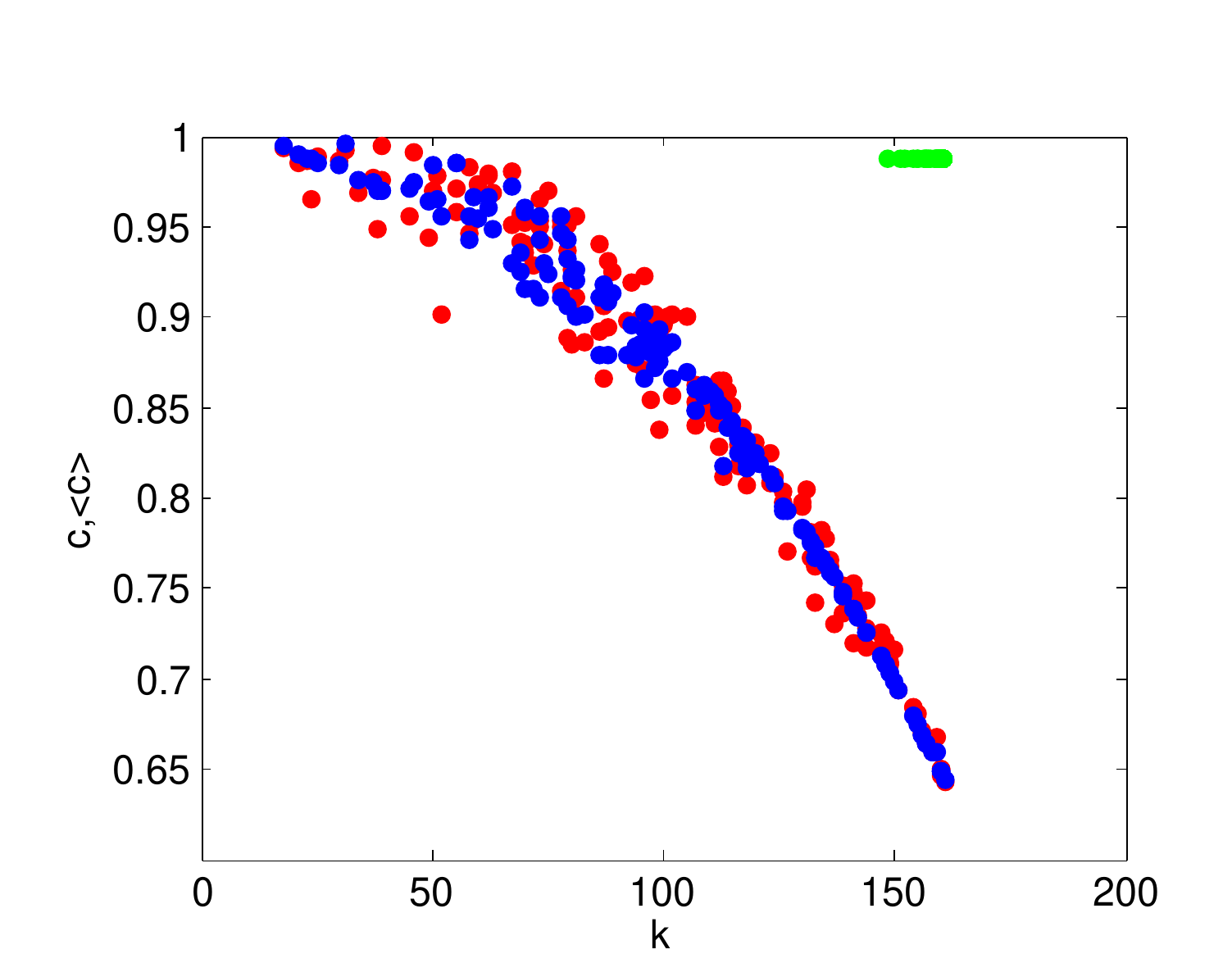}}
{\includegraphics[width=0.4\textwidth]{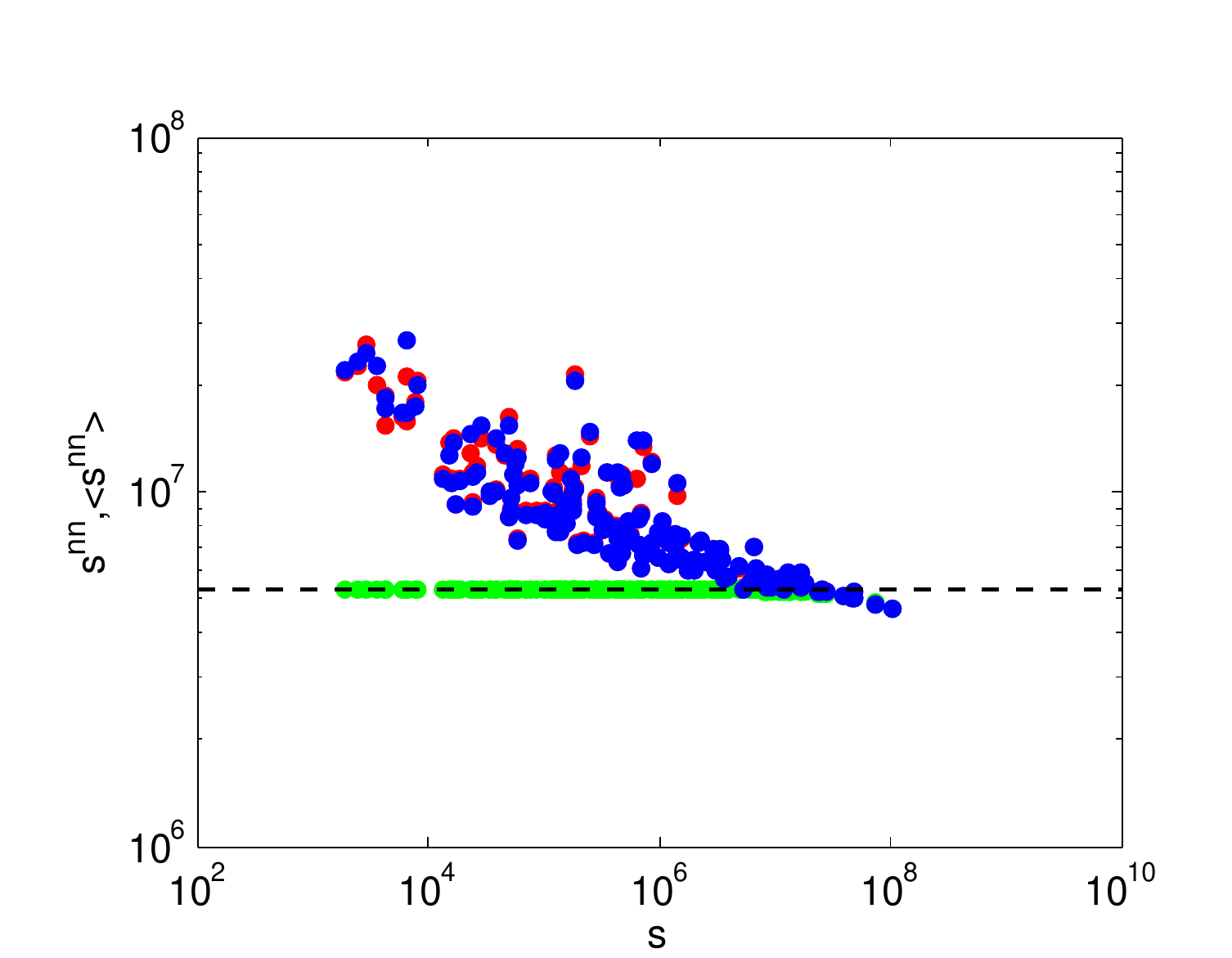}}
\hspace{-5mm}
 {\includegraphics[width=0.4\textwidth]{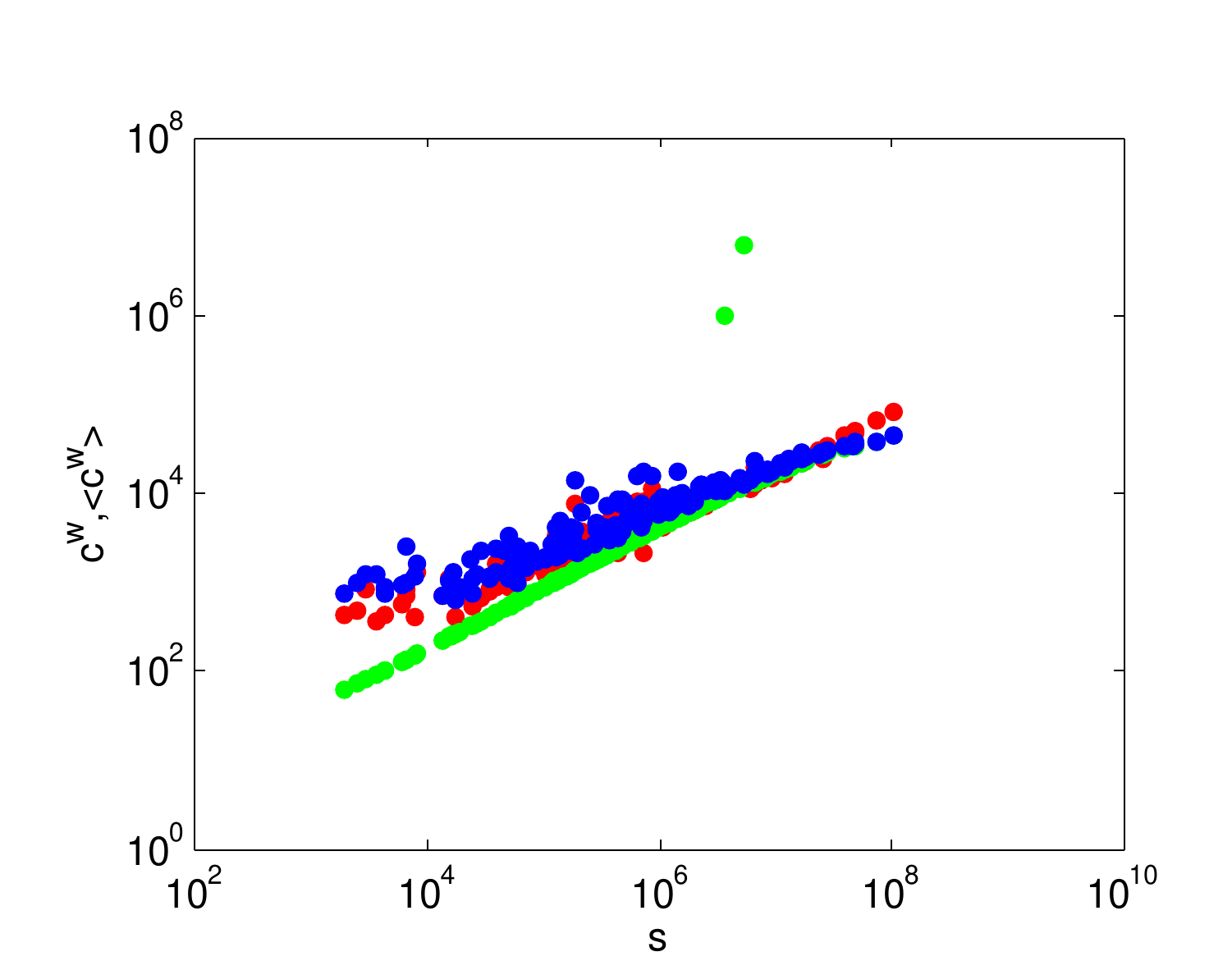}}
\caption{Comparison between the observed undirected binary and weighted properties (red points) and the corresponding ensemble averages of the WCM (green points) and the ECM (blue points) for the aggregated WTW in the 2002 snapshot. Top left: Average Nearest Neighbor Degree $k_i^{nn}$ versus degree $k_i$. Top right: Binary Clustering Coefficient $c_i$ versus degree $k_i$. Bottom left: Average Nearest Neighbor Strength $s_i^{nn}$ versus strength $s_i$. Bottom right: Weighted Clustering Coefficient $c^w_i$ versus strength $s_i$. }\label{AggBUN}
\end{figure*}

In fig. \ref{AggBUN} we show, for the snapshot in year 2002, the scatter plots between node degree ($k_i$) and higher-order binary properties  ($k^{nn}_i, c_i$), as well as those between node strength ($s_i$) and higher-order weighted properties ($s^{nn}_i, c^w_i$). 
From an economic point of view, $k^{nn}_i$ and $c_i$ give information about indirect interactions over paths of length $2$ and $3$ respectively, since terms of the form $a_{ij}a_{jk}$ and $a_{ij}a_{jk}a_{ki}$ are involved in the definitions of these quantities. In accordance with the existing literature, we find decreasing trends of both $k_i^{nn}$ and $c_i$ versus $k_i$. 
This confirms that it is very likely to find nodes with many trade partners connected to nodes with small degree (disassortativity), while trade partners of poorly connected nodes are highly inter-connected. 
Similar considerations hold true when we consider edge weights, as $s^{nn}_i$ and $c^w_i$ involve indirect paths of length $2$ and $3$ respectively,  now including mixed information about topology and weights. 
Intensively trading countries are found to be connected with poorly trading countries, confirming a dissasortative pattern (even if less prominent than in the binary case) at a weighted level. 

Besides the observed values of the aforementioned quantities, in fig. \ref{AggBUN} we also plot the corresponding expected values predicted by the ECM, as well as those predicted by the WCM.
The latter represent the starting point of our analysis, because they are intended to merely replicate the results in ref. \cite{Squartini_etal_2011b_pre}. 
Indeed, we confirm that the WCM is in striking disagreement with the observed values.
While, at a binary level, the empirical degree correlations and clustering structure of the ITN are excellently reproduced by the BCM (which uses only the knowledge of the degree sequence) \cite{Squartini_etal_2011a_pre}, at a weighted level the observed network properties are very different from the predictions of the WCM (which, naively, is the obvious extension of the BCM to weighted graphs) \cite{Squartini_etal_2011b_pre}. These results are robust over time and for various resolutions (i.e., for different levels of aggregation of traded commodities). 

It is important to realize the origin of the disagreement between the WCM and the real network.
We note that the expected values under the WCM are similar to those predicted for a fully connected topology. Indeed, for a complete network we have
\begin{eqnarray}
\langle k_i\rangle &=& N-1 \label{1} \\
\langle k^{nn}_i \rangle &=& N-1 \label{2} \\
\langle c \rangle &=& 1 \label{3} \\
\langle s^{nn}_i \rangle & =& \frac{\sum_i s_i}{N-1} = \frac{2W_{TOT}}{N-1} 
\label{4}
\end{eqnarray}
where $N$ stands for the number of nodes in the network, while $W_{TOT}$ is the total weight of all edges. The above predictions can be confirmed in fig. \ref{AggBUN}.
So, the main reason why the WCM fails is the fact that it generates unrealistically dense (and sometimes almost fully connected) networks  \cite{Squartini_etal_2011b_pre,mastrandrea2013enhanced}.
We also note that, despite the apparent good agreement between the observed weighted clustering coefficient and its expected value under the WCM (fig. \ref{AggBUN}), one can show that the empirical total level of clustering is in general higher than the one predicted by WCM, both throughout the temporal evolution of the system and across its commodity-specific layers \cite{Squartini_etal_2011b_pre}.

We note that another study \cite{fronczak2012statistical} also applied a variant of the WCM (basically assuming non-negative but real-valued, instead of integer-valued, edge weights) to the WTW.
However, the quantities used therein to test the model against the data did not depend in any way on the adjacency matrix $A$, i.e. they were entirely topology-independent. 
As a result, the authors concluded that the observed WTW is a typical member of the WCM. Our analysis, together with ref. \cite{Squartini_etal_2011b_pre}, shows that monitoring more (topological) properties leads to the opposite conclusion.
Indeed, it is easy to show that the variant of the WCM used in ref. \cite{fronczak2012statistical} predicts a rigorously fully connected network, precisely because the assumption of real-valued weights implies a zero probability of missing links (zero weights). 
Therefore the real-valued WCM encounters the limitations discussed above in an even more extreme way.

We now come to the predictions of the ECM shown in fig. \ref{AggBUN}. 
In marked contrast with the WCM, the ECM performs very well and reproduces both the binary and weighted properties of the WTW.
Firstly, we find a definitely improved agreement for the binary trends ($k^{nn}_i$ and $c_i$ versus $k_i$): the ECM shows an expected trend that follows the data very closely. 
For these binary properties, the predictions of the ECM are even closer to the observed cloud of points than the monotonic curves predicted by the BCM for the same network, as a visual comparison with the results shown in ref. \cite{Squartini_etal_2011a_pre} immediately reveals. Secondly, we also find a significantly better agreement, with respect to the WCM, between the observed and the randomized weighted trends ($s^{nn}_i$ and $c^w_i$ versus $s_i$).

These results imply that the knowledge of both the number of trade partners of each node and the total amount of trade flowing through each country is highly informative about the higher-order and non local dynamics of the whole network.
More in general, the local binary information is crucial in order to predict the weighted structure itself. In turn, this means that the weighted information alone does not allow a deep understanding of the topology of the network.
We therefore confirm our recent finding  \cite{mastrandrea2013enhanced} that the na\"ive expectation that weighted quantities are \emph{per se} more informative than the corresponding binary ones is fundamentally incorrect. 
To further validate these findings, in what follows we explore the evolution of the same properties over time, and across different layers of the trade multiplex.

\begin{figure*}[t]
\centering
{\includegraphics[width=.8\textwidth]{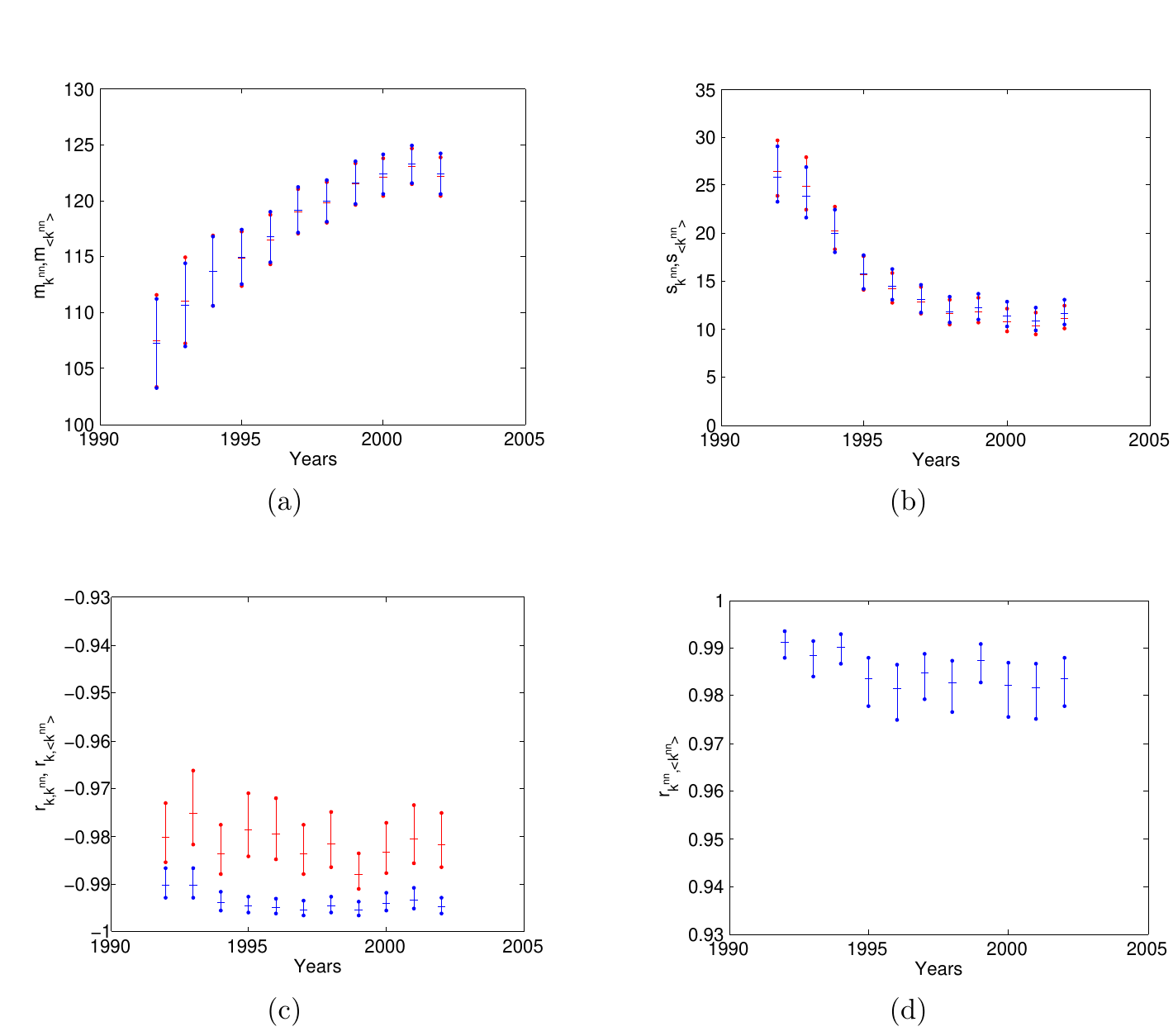}}
\caption{Temporal evolution of the  Average Nearest Neighbour Degree ($k^{nn}_i$) from 1992 to 2002 and comparison with the corresponding maximum-entropy ensembles with specified degrees and strengths (ECM). (a) average of $k^{nn}_i$ across all vertices (red: obs., blue: randomized); (b) standard deviation of $k^{nn}_i$ across all vertices; (c) correlation coefficient between $k^{nn}_i$ and $k_i$; (d) correlation coefficient between $k^{nn}_i$ and $\langle k^{nn}_i \rangle$. Red: observed values; blue: expected values. The $95\%$ confidence intervals of all quantities are represented as vertical bars.}\label{knn}
\end{figure*} 

\begin{figure*}[t]
\centering
{\includegraphics[width=.8\textwidth]{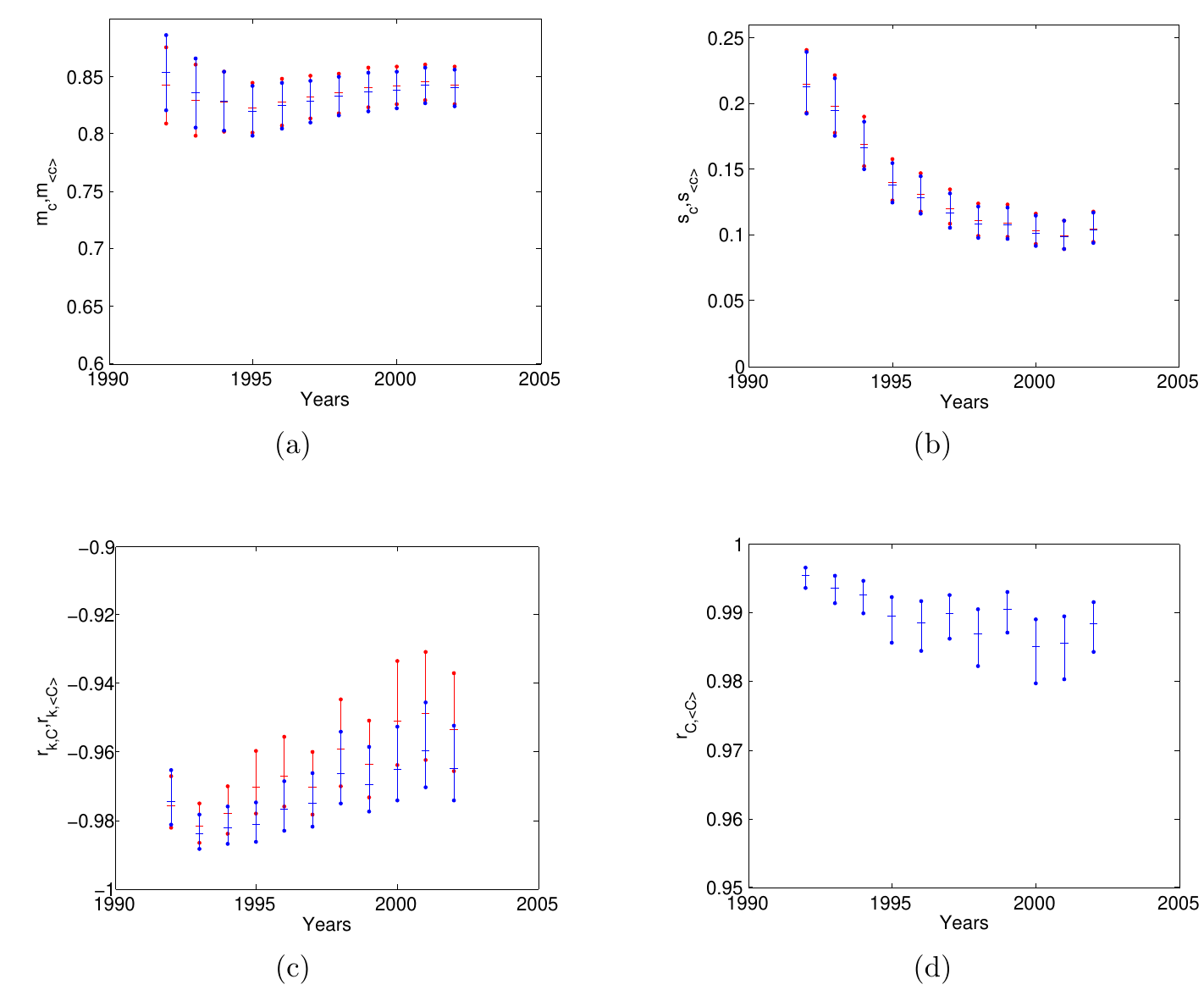}}
\caption{Temporal evolution of the Binary Clustering Coefficient ($c_i$) from 1992 to 2002, and comparison with the corresponding maximum-entropy ensembles with specified degrees ans strengths (ECM). (a) average of $c_i$ across all vertices (red: obs., blue: randomized); (b) standard deviation of $c_i$ across all vertices;
(c) correlation coefficient between $c_i$ and $k_i$; (d) correlation coefficient between $c_i$ and $\langle c_i \rangle$. Red: observed values; blue: expected values. The $95\%$ confidence intervals of all quantities are
represented as vertical bars.}\label{cc}
\end{figure*}

\begin{figure*}[t]
\centering
{\includegraphics[width=.8\textwidth]{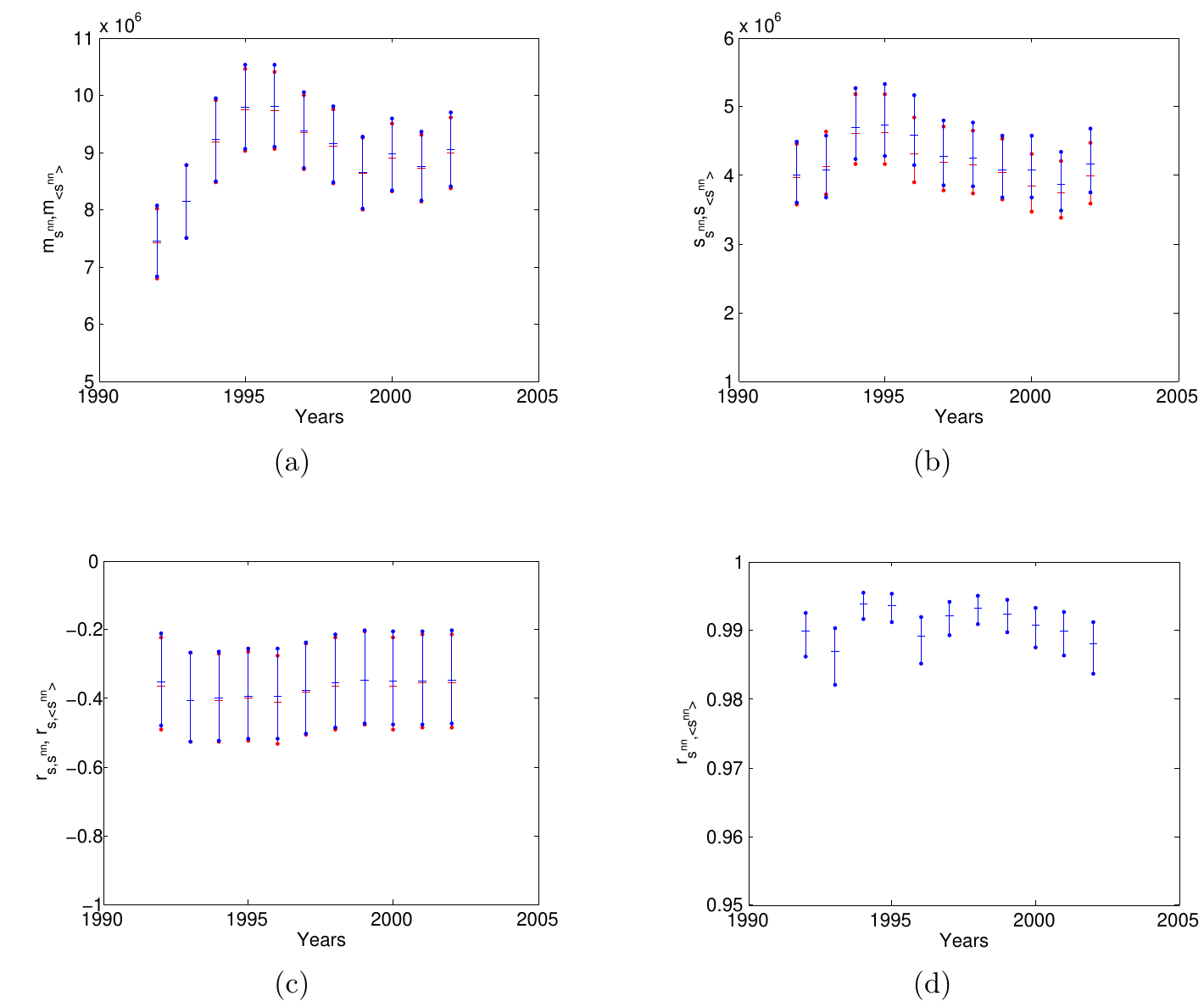}}
\caption{Temporal evolution of the properties of the ANNS $s^{nn}_i$ in the 1992-2002 snapshots of the observed undirected WTW and of the corresponding maximum-entropy ensembles with specified degrees ans strengths: (a) average of $s^{nn}_i$ across all vertices (red: obs., blue: randomized); (b) standard deviation of $s^{nn}_i$ across all vertices; (c) correlation coefficient between $s^{nn}_i$ and $s_i$; (d) correlation coefficient between $s^{nn}_i$ and $\langle s^{nn}_i \rangle$. Red points stands for observed values, blue for the randomized ones; the $95\%$ confidence intervals of all quantities are
represented as vertical bars.}\label{snn}
\end{figure*}

\begin{figure*}[t]
\centering
{\includegraphics[width=.8\textwidth]{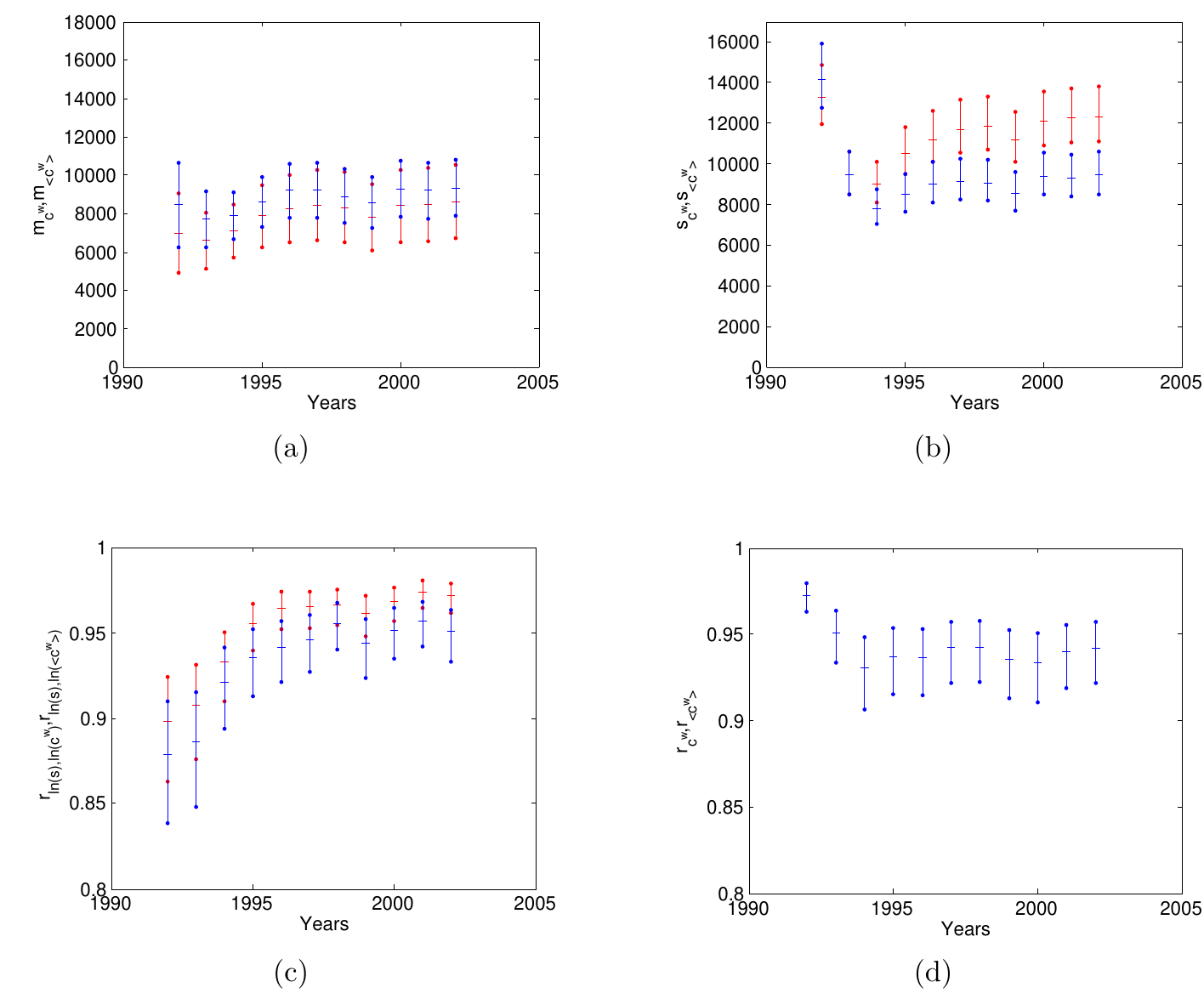}}
\caption{Temporal evolution of the properties of the WCC $c^w_i$ in the 1992-2002 snapshots of the observed undirected WTW and of the corresponding maximum-entropy ensembles with specified degrees ans strengths: (a) average of $c^w_i$ across all vertices (red: obs., blue: randomized); (b) standard deviation of $c^w_i$ across all vertices;
(c) correlation coefficient between $c^w_i$ and $s_i$; (d) correlation coefficient between $c^w_i$ and $\langle c^w_i \rangle$. Red points stands for observed values, blue for the randomized ones; the $95\%$ confidence intervals of all quantities are represented as vertical bars.}\label{ccww}
\end{figure*}

\subsection{Temporal evolution\label{sec:time}}
We first test the robustness of our results over time, by replicating the previous analysis on the 11 yearly snapshots of the aggregated network. We measure the same properties, as well as their expected values, as defined in sec. \ref{wtwagg}, where now the matrices $A$ and $W$ refer to each of the various snapshots of the system, i.e. $w_{ij}\equiv w_{ij}^{AGG}(t)$ for $t=1992,1993, \dots 2002$.

We show our results in the following, more compact way. 
We consider the four network properties defined in eqs. \eqref{ANND}-\eqref{WCC} separately, and for each network property we take the list of observed values (e.g. $\{k^{nn}_i\}$) and the list of  expected values under the ECM (e.g. $\{ \langle k^{nn}_i \rangle \}$). For each of the two lists, we compute the following three metrics: 
\emph{i)} the average value of the list; \emph{ii)} the standard deviation of the list; \emph{iii)} the Pearson correlation coefficient between the list and the list of `natural' constraints that we used above as the independent variable in the relevant scatter plot (e.g. $\{k_i\}$). 
As a fourth metric, we also calculate: \emph{iv)} the Pearson correlation coefficient between the list of expected values and the list of observed values (e.g. between $\{k^{nn}_i\}$ and $\{ \langle k^{nn}_i \rangle \}$).
Each of these four metrics summarizes one aspect of the scatter plot (of the type shown in fig. \ref{AggBUN}) for the structural property under consideration, thereby allowing us to compactly track the evolution of the system over time.

It should be noted that the use of the correlation coefficient (\emph{iv}) is more appropriate than that of the coefficient (\emph{iii}), since a perfect agreement between model and data implies an equality between expected and observed properties. 
Such an equality is a proper form of linear correlation, for which we expect the Pearson correlation coefficient to achieve its maximum value 1.
In case of partial agreement, a value below 1 correctly indicates a lack of equality (i.e. a lack of linearity) between expected and observed values. 
By contrast, as clear from fig. \ref{AggBUN}, there is a \emph{nonlinear} correlation between higher-order properties (e.g. $k^{nn}_i$) and the chosen constraints (e.g. $k_i$).
Therefore we do not expect the correlation coefficient (\emph{iii}) to be in general close to $\pm 1$: rather, we merely expect that a perfect agreement between model and data leads to similar values of the coefficients derived from expected and observed values.
However, in case of partial agreement we can no longer expect a consistency between the two, since the nonlinear character of both observed and expected trends might have an uncontrolled effect on the linear correlation coefficient. 
Thus our choice of including the correlation coefficient (\emph{iii}) is mainly due to the need of comparing our results with previous studies, as we now discuss.

The analysis described above is shown in four figures (one for each structural property) of four panels each (one for each metric).
Specifically, figs. \ref{knn}, \ref{cc}, \ref{snn} and \ref{ccww} show the temporal evolution (time series) of the four metrics mentioned above, for the Average Nearest Neighbour Degree, Binary Clustering Coefficient, Average Nearest Neighbour Strength and Weighted Clustering Coefficient respectively.
For each property and each point in time, we also plot the associated $95\%$ confidence intervals. 
Again, this kind of visualization coincides intentionally with that used in previous analyses of the same data, where the BCM \cite{Squartini_etal_2011a_pre} and WCM \cite{Squartini_etal_2011b_pre} were used.
Our use of the same four metrics and the same four properties allows us to compare the performance of the ECM, i.e. of the combination of degrees and strengths, with that of the other two null models where only one constraint is used.

Figures \ref{knn} and \ref{cc} show that the time series of both average value and standard deviation of the Average Nearest Neighbor Degree and Clustering Coefficient are all perfectly replicated by the ECM predictions over time.
The correlation coefficient (\emph{iii}) of both $k_{i}^{nn}$ and $c_i$ with $k_i$ is the only metric where there is some minor disagreement, however this cannot be interpreted as a statistically significant deviation, as we discussed. 
The tight agreement between observed and expected values over time is best confirmed by the correlation coefficient (\emph{iv}), which is always very close to $1$. 
All these results are perfectly in line with what is obtained using the BCM, i.e. when only the degree is enforced as a constraint, on the purely binary representation of the same data  \cite{Squartini_etal_2011a_pre}. 
This means that, by simultaneously preserving degrees and strengths, the ECM does not diminish the ability of the BCM to predict the binary topology of the WTW (as we have seen in fig. \ref{AggBUN}, the ECM actually \emph{improves} the already good fit of the BCM to the data).

Figures \ref{snn} and \ref{ccww} show that also for the weighted network properties there is an excellent agreement between the observed values and the corresponding expectations over the ECM, for the whole period.
The ECM is able to accurately reproduce the temporal trends of the average of both $s^{nn}_i$ and $c^w_i$, as well as their standard deviation.
The correlation (\emph{iii}) of these properties with node strength is also well replicated by the ECM in the whole period. 
Finally, the correlation coefficient  (\emph{iv}) between observed and randomized properties is almost 1 all the time.
These results are very different from what is obtained using the WCM on the same data \cite{Squartini_etal_2011b_pre}. 
Again, the WCM is completely unable to replicate the observed trends.
The addition of purely binary information, embodied in the number of node partners, makes the ECM very powerful in predicting the higher-order properties of the WTW, throughout the temporal window considered.

\subsection{Information-theoretic comparison\\ of the WCM and the ECM\label{sec:AIC}}

Before proceeding to the analysis of individual layers of the trade multiplex, we perform an important check of the statistical appropriateness of the results obtained so far.
This check is needed for the following reason.
It is obvious that, by including more constraints, the ECM achieves a better fit than the WCM. 
However, in principle the gain in accuracy (better fit) might be smaller than the loss in parsimony (more parameters), i.e. the ECM might overfit the network.
To rigorously make this assessment, we perform an information-theoretic comparison of the two models in terms of the achieved trade-off between accuracy and parsimony. 

Information-theoretic criteria exist \cite{burnham2004multimodel} to assess whether the increased accuracy of a model with more parameters comes at the price of an excessive loss of parsimony. The most popular choice is the Akaike's Information Criterion ($AIC$), showing that the optimal trade-off between accuracy and parsimony is achieved by discounting the number of free parameters from the maximized log-likelihood \cite{burnham2004multimodel}. 

To compute $AIC$, we therefore first need to calculate the maximized log-likelihood of the two models.
As we mentioned, the WCM can be obtained as a particular case of the ECM by setting $x_i=1$ for all $ i$, i.e. by `switching off' the Lagrange parameters controlling for the degrees. The log-likelihood of the WCM is therefore the reduced function $\mathcal{L}(\vec{1},\vec{y})$ of $N$ variables, and is maximized by a new vector $\vec{y}^{**} \ne \vec{y}^*$, where $(\vec{x}^*,\vec{y}^*)$ stands for the solution of the ECM and $(\vec{1},\vec{y}^{**})$ for the solution of the WCM for the same observed network. 

Given the maximized log-likelihood of our two competing models, we calculate the size-corrected \cite{burnham2004multimodel} version of $AIC$, denoted as $AIC_c$, as follows:
\begin{equation}
AIC_c^{ECM}\equiv -2\mathcal{L}(\vec{x}^*,\vec{y}^*)+4N+ \frac{8N(2N+1)}{N^2-5N-2}
\end{equation}
\begin{equation}
AIC_c^{WCM}\equiv -2\mathcal{L}(\vec{1},\vec{y}^{**})+2N+\frac{4N(N+1)}{N^2-3N-2}
\end{equation}
The last term on the r.h.s. of both equations provides the correction to $AIC$ when the number of parameters is not negligible with respect to the sample cardinality (as a rule of thumb, when $n/k<40$, $n$ being the cardinality of the sample and $k$ being the number of parameters \cite{burnham2004multimodel}), thus further reducing the probability of overfitting. Notice that, when $n\gg k^2$, the additional term converges to $0$, recovering the standard form of $AIC$ for the ECM \cite{mastrandrea2013enhanced}. 
Precisely for this reason, $AIC_c$ should be always employed regardless of the value of $n/k$ \cite{burnham2004multimodel}.

The model that achieves the best trade-off between accuracy and parsimony is the one with the smallest value of $AIC_c$. However, if the difference of the $AIC_c$ values is small, the two models will still be comparable. 
A quantitative criterion to statistically interpret the differences of $AIC_c$ is given by the so-called Akaike Weights, which quantify the weight of evidence in favour of a model, i.e., the probability that the model is the best one among the (two) models considered. 
In our case, these weights read
\begin{eqnarray}
w_{AIC_c}^{ECM}&\equiv&\frac{e^{-AIC_c^{ECM}/2}}{e^{-AIC_c^{ECM}/2}+e^{-AIC_c^{WCM}/2}}\\
\label{eq_wmcm}
w_{AIC_c}^{WCM}&\equiv&1-w_{AIC_c}^{ECM}.
\label{eq_wwcm}
\end{eqnarray}

Given a real network, a low value of $w_{AICc}^{ECM}$ will indicate that the addition of the degree sequence is redundant (the relevant local constraints effectively reduce to the strength sequence, so the standard WCM is preferable), while a high value of $w_{AICc}^{ECM}$ will indicate that, in addition to the strength sequence, the degrees must be separately specified. 
We stress that the result of this procedure is not predictable a priori (it depends on the numerical values of $\{s_i\}$ and $\{k_i\}$) and can only be achieved after a comparison of the two model on the specific data at hand.

In Table \ref{aictab} we show the results for the two competing models, for the particular year 2002. We also used the Bayesian Information Criterion ($BIC$), \cite{burnham2004multimodel}, which puts a higher penalty on the number of parameters. 
Both criteria yield (up to machine precision) a unit probability that the ECM is the best model. 
This confirms that the addition of the degree sequence as a constraint is non-redundant and extremely informative for the prediction of the WTW properties.
We systematically found the same result for all temporal snapshots considered in sec. \ref{sec:time} and all commodity classes that will be illustrated in sec. \ref{sec:multiplex} (values not shown for brevity).

The above finding implies that the world trade multiplex is yet another system consistent with the `irreducibility conjecture' we proposed in ref. \cite{mastrandrea2013enhanced}. This conjecture states that, in real-world weighted networks, the strengths are not necessarily more informative than the degrees; rather, they are a complementary piece of information.
Strengths and degrees are therefore `irreducible' to each other, because they constrain the network in fundamentally different ways. 

\begin{table}[b]
\centering
\begin{tabular}{ccccc}
\hline
\hline
  & $AIC_c$ & $ BIC$  &  $ w_{AIC_c}$ & $ w_{BIC}$ \\
\hline
$\bf WCM$ & $209,972$ & $211,179$ & $0$ & $0$\\
$\bf ECM$ & $165,731$ & $168,137$ & $1$ & $1$\\
\hline
\end{tabular}
\caption{$AIC_c$ and $BIC$ values, along with the associated $AIC_c$ and $BIC$ weights, for the two null models (WCM and ECM) applied to the WTW in 2002.}
\label{aictab}
\end{table}

An important macroeconomic implication is that any model aimed at reproducing the WTW (either statically, over time, and/or across its layers) should not discard any of the two quantities. This conclusion sets an important challenge for future models of trade, given that most models in the literature, and most notably Gravity Models \cite{duenas2011modeling,myTinbergen}, mainly focus on weighted properties (trade volumes) and largely discard purely topological information.

\begin{figure*}[t]
\centering
\includegraphics[width=\textwidth]{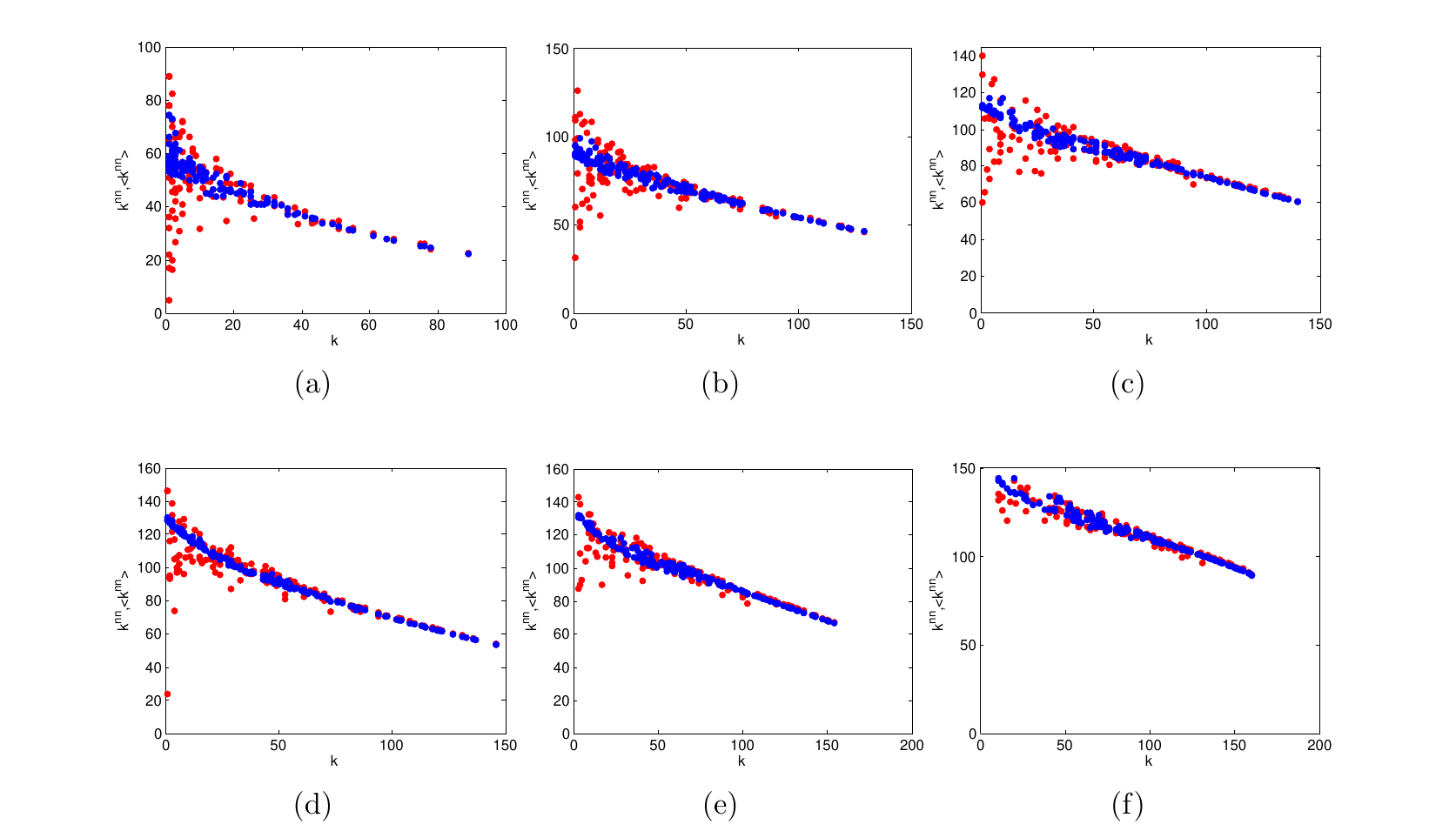}
\caption{Average Nearest Neighbour Degree ($k^{nn}_i$) versus node degree ($k_i$) in the 2002 snapshots of the commodity-specific (disaggregated) versions of the observed binary undirected WTW (red points), and corresponding average over the maximum entropy ensemble with specified degrees and strengths (blue points): a) commodity $93$; b) commodity $09$; c) commodity $39$; d) commodity $90$; e) commodity $84$; f) aggregation of the top $14$ commodities (see table \ref{tab} for details). From a) to f), the intensity of trade and level of aggregation increases.}\label{knnCom}
\end{figure*}

\begin{figure*}[t]
\centering
\includegraphics[width=\textwidth]{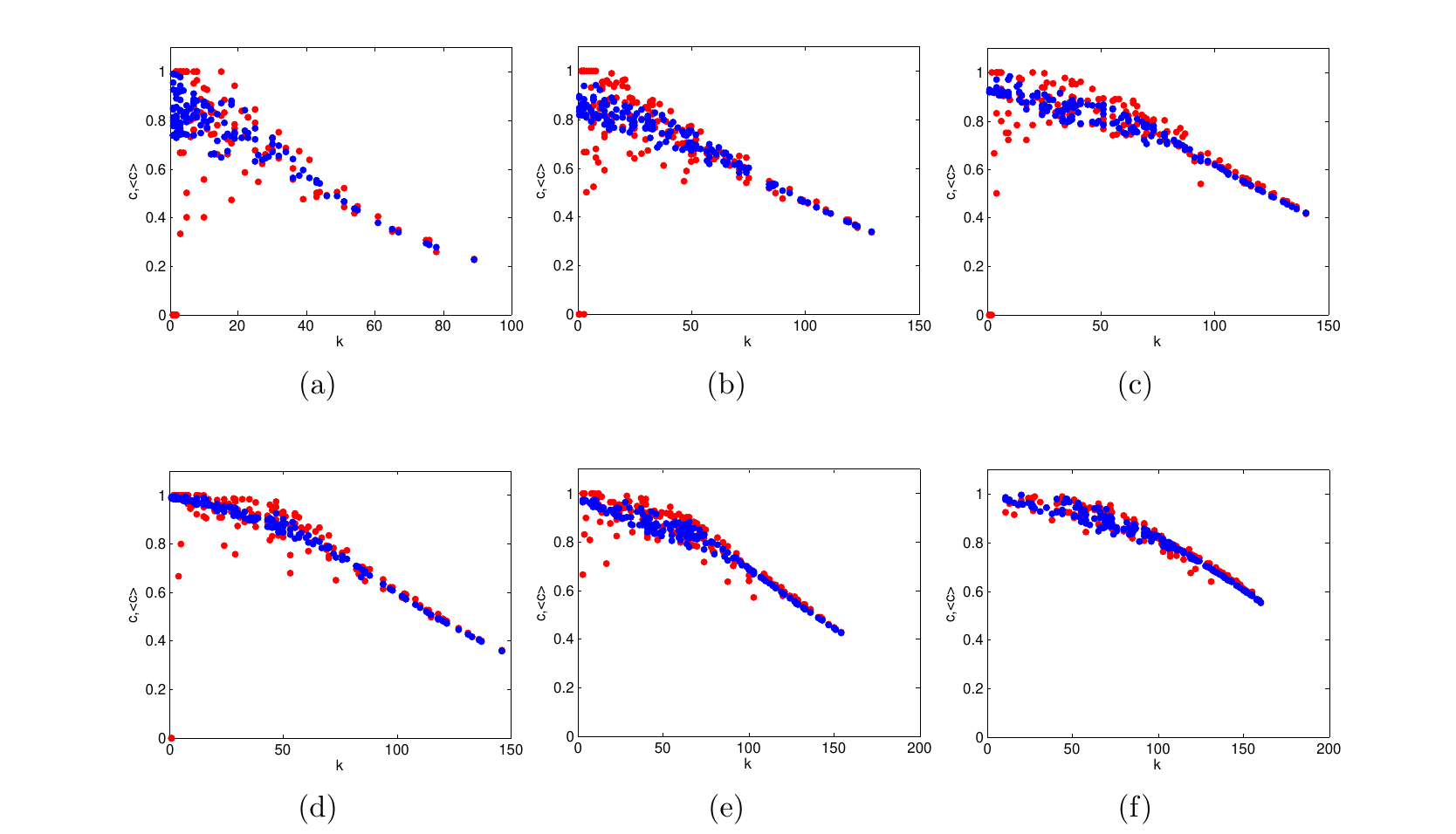}
\caption{Binary Clustering Coefficient ($c_i$) versus node degree ($k_i$) in the 2002 snapshots of the commodity-specific (disaggregated) versions of the observed binary undirected WTW (red points), and corresponding average over the maximum entropy ensemble with specified degrees and strengths (blue points): a) commodity $93$; b) commodity $09$; c) commodity $39$; d) commodity $90$; e) commodity $84$; f) aggregation of the top $14$ commodities (see table \ref{tab} for details). From a) to f), the intensity of trade and level of aggregation increases.}\label{ccCom}
\end{figure*}

\subsection{Multiplex analysis:\\ commodity-specific trade networks\label{sec:multiplex}}

We conclude our empirical analysis by studying the individual networks formed by imports and exports of single (classes of) commodities.

As we described in sec. \ref{sec:data1}, our data set resolves the trade multiplex into $C=97$ layers. 
While it is unfeasible to replicate the analysis described so far on each of the $C\times T=97\times 11=1067$ networks resulting from the evolution of the $C$ layers over $T$ years, we selected the same subset of layers as in refs. \cite{Squartini_etal_2011a_pre,Squartini_etal_2011b_pre}.
This choice allows us to gain information about the performance of the ECM on layers with a broad range of sparseness and edge weight: while the aggregate WTW considered above is a highly dense network (with density around $0.5$) with large total weight, the commodities we selected vary considerably in their intensity and level of connectivity. 
The selection includes the two least traded commodities (in terms of total trade value, i.e. total edge weight) in the entire data set (`Arms and ammunition', $c = 93$, and `Coffee, tea, mate \& spices', $c=9$), two intermediate ones (`Plastics', $c = 39$, and `Optical, photographic, cinematographic, measuring, checking, precision, medical or surgical instruments', $c=90$), the most traded one (`Nuclear reactors, boilers, machinery and mechanical appliances', $c = 84$), plus the network formed by combining all the top $14$ commodities described in sec.\ref{sec:data1} together (see Table \ref{tab} for details). The last sub-network represents an intermediate level of aggregation between single commodities and the completely aggregated data analysed in sec. \ref{wtwagg}. 
The six (classes of) commodities described above, plus the fully aggregated network itself, form a set of seven (combinations of) layers in increasing order of trade intensity, link density, and aggregation.

\begin{figure*}[t]
\centering
\includegraphics[width=\textwidth]{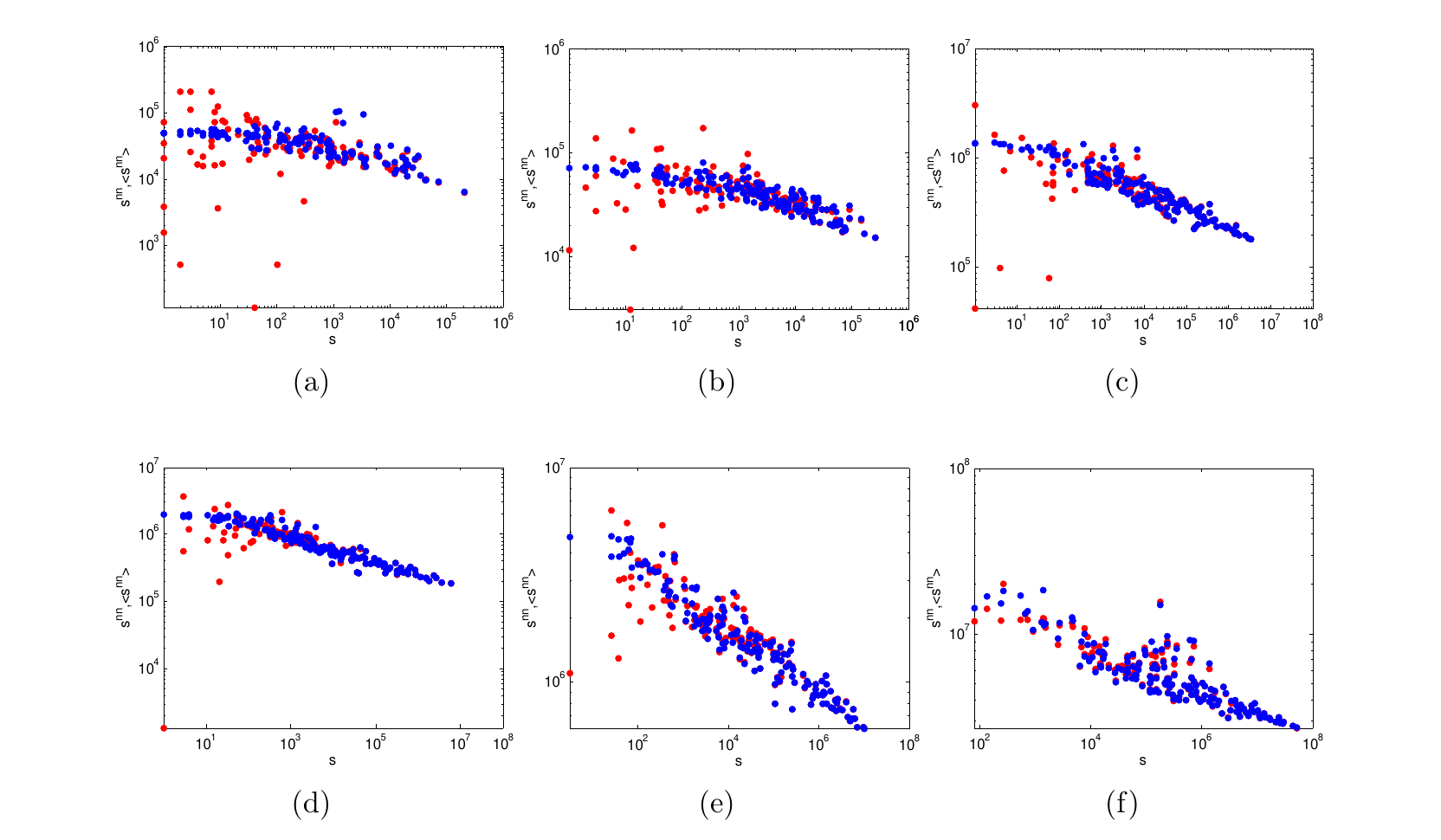}
\caption{Average Nearest Neighbour Strength ($s^{nn}_i$) versus node strength ($s_i$) in the 2002 snapshots of the commodity-specific (disaggregated) versions of the observed binary undirected WTW (red points), and corresponding average over the maximum entropy ensemble with specified degrees and strengths (blue points): a) commodity $93$; b) commodity $09$; c) commodity $39$; d) commodity $90$; e) commodity $84$; f) aggregation of the top $14$ commodities (see table \ref{tab} for details). From a) to f), the intensity of trade and level of aggregation increases.}\label{snnCom}
\end{figure*}

\begin{figure*}[t]
\centering
\includegraphics[width=\textwidth]{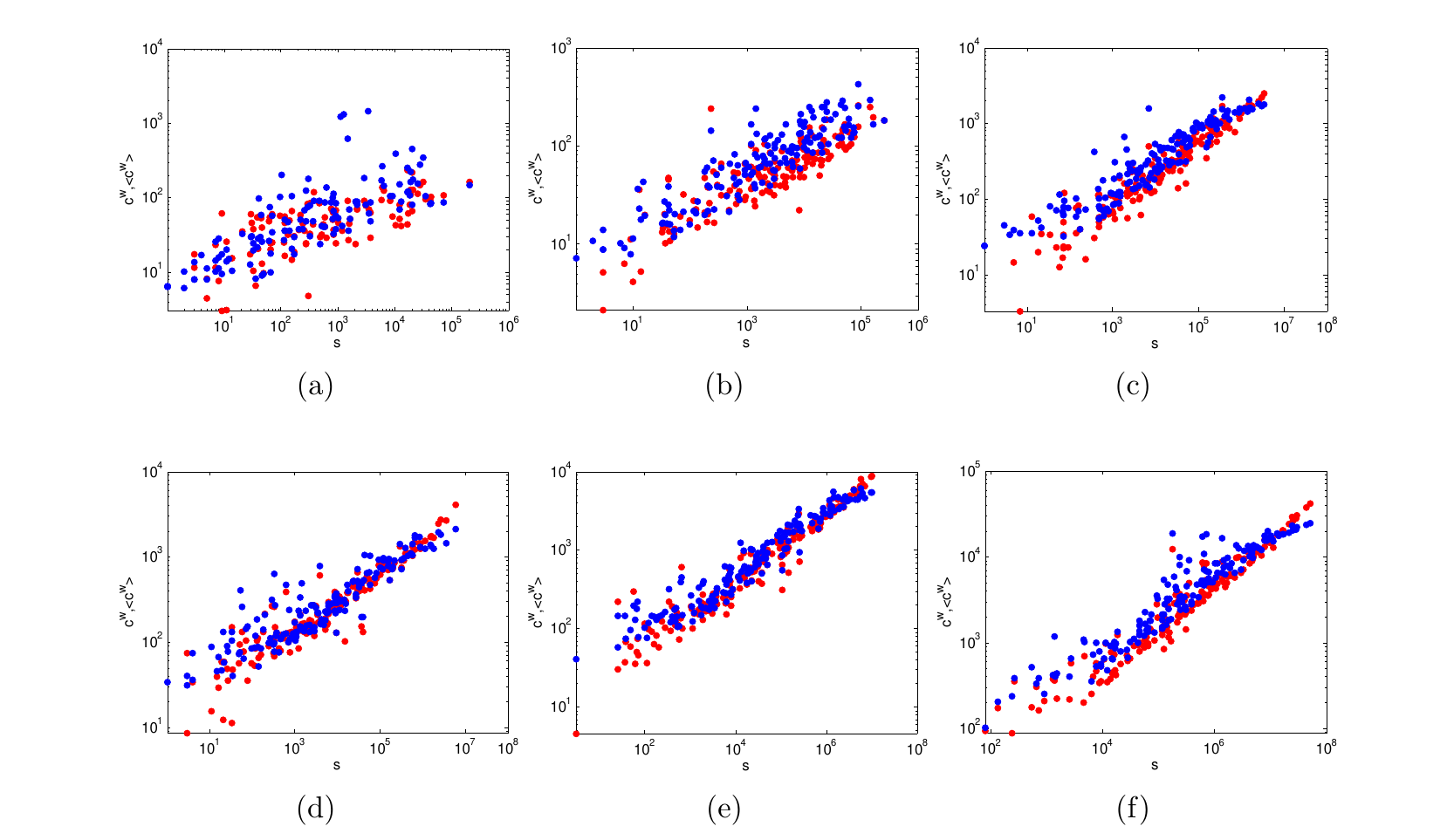}
\caption{Weighted Clustering Coefficient ($c^w_i$) versus node strength ($s_i$) in the 2002 snapshots of the commodity-specific (disaggregated) versions of the observed binary undirected WTW (red points), and corresponding average over the maximum entropy ensemble with specified degrees and strengths (blue points): a) commodity $93$; b) commodity $09$; c) commodity $39$; d) commodity $90$; e) commodity $84$; f) aggregation of the top $14$ commodities (see table \ref{tab} for details). From a) to f), the intensity of trade and level of aggregation increases.}\label{wccCom}
\end{figure*}

We consider the scatter plots of both binary and weighted higher-order properties for the 2002 snapshot of the above layers, as we did in ref.\ref{wtwagg} for the aggregate network.
This is shown in figs. \ref{knnCom}, \ref{ccCom}, \ref{snnCom} and \ref{wccCom} for the Average Nearest Neighbour Degree, Binary Clustering Coefficient, Average Nearest Neighbour Strength and Weighted Clustering Coefficient respectively.
Remarkably, we find that the results obtained in our aggregated analysis also hold for individual commodities, independently of the level of aggregation. 
Also for the temporal evolution and information-theoretic analysis of the system, our results are very similar to those found for the aggregate network in secs. \ref{sec:time} and \ref{sec:AIC} respectively, but are not shown here for the sake of brevity.

The binary results confirm, and slightly improve, the performance of the BCM on the same data \cite{Squartini_etal_2011a_pre}.
However, the excellent agreement between observed and randomized \emph{weighted} properties in the commodity-specific case is more surprising and represents a marked improvement with respect to the predictions of the WCM for the same system \cite{Squartini_etal_2011b_pre}.
The case of the weighted clustering coefficient is particularly interesting in this sense. 
Indeed, while for the aggregate network the WCM gives a reasonable prediction of (only) this quantity (see fig. \ref{AggBUN}), this outcome is not robust to disaggregation: individual layers of the empirical multiplex show a deviation from the WCM, which incresases with the sparseness of the layer \cite{Squartini_etal_2011b_pre}. 
On the contrary, fig. \ref{wccCom} shows that the ECM accurately replicates the observed clustering, as well as all the other properties under investigation, for every level of disaggregation.

In general, we observe a slightly worse agreement in layers with smaller density/volume (this is especially true for the weighted properties).
However, this appears to be mainly due to the fact that the empirical scatter plots associated with less traded goods are more dispersed (due to less statistics).
As expected, this effect is even more pronounced for the weighted quantities, because the range of allowed values for the strength is wider than that for the degree.


\section{Discussion \label{sec:discussion}}
We now discuss our result in the light of various lines of research in economics.
In particular, we focus on the role of local properties in economic networks and on the relation between our results and the established knowledged about intensive and extensive margins of trade. 

\subsection{The role of local properties in economic networks}
In economic and financial networks, the total strength of the connections reaching a node  has generally an important meaning, such as the size of supply and demand, import and export, or financial exposure. 
Hence, generating random ensembles of networks matching the observed strengths of all nodes is crucial in order to detect interesting deviations of a known empirical network from economically meaningful benchmarks, to reconstruct the most likely structure of an unknown network from purely local information, or finally to define a model of economic networks specified by node-specific properties. 

Our results show that, in order to correctly reproduce the whole structure of the WTW as a weighted network, the degree sequence must be constrained in addition to the strength sequence. 
From the general point of view of network reconstruction, these findings consolidate and widely extend the results in \cite{mastrandrea2013enhanced}. 
We confirmed the effectiveness of the ECM in reproducing the higher-order properties of the WTW starting from local constraints, and succesfully tested the robustness of the model with respect to several temporal snapshots and levels of aggregation. 
So, while the strength sequence (a weighted constraint) turns out to be uninformative about the binary topology of the WTW, the degree sequence (a binary constraint) plays a fundamental role in reproducing its weighted structure. 
This highly asymmetric role of binary and weighted constraints is a non-trivial result.

From an economic perspective, the fact that purely local information is enough in order to reproduce the large-scale structure of the WTW implies that parsimonious  models of international trade can largely discard additional mechanisms besides those accounting for the number of partners and total trade of world countries. 
The importance of reproducing and/or explaining the degrees of all world countries, first pointed out in \cite{Squartini_etal_2011a_pre}, is confirmed by our study, and shown to hold even when one considers the weighted representation of the WTW.
This strengthens the view that theories and models of trade, when aiming at explaining the international network structure, should seriously focus on the number of trade partners of countries as an important target quantity to replicate.

Of course, the above considerations leave an important point open, namely the role played by other properties of (expected) economic relevance in shaping the structure of the international trade network. For instance, Gravity Models \cite{Bergstrand1985,chaney2008distorted,duenas2011modeling,myTinbergen} predict that both country-specific (mainly the GDP) and dyadic quantities, such as geographic distance (which is a proxy of trade resistance) and other extra factors (such as common currency, common language, borderding conditions, etc.), do play an important role.
The GDP is directly related (and roughly proportional) to the total trade of a country, i.e. its strength \cite{Fagiolo2008physa,Fagiolo2009jee}. It is also related to the number of trade partners, i.e. the degree, in a highly non-linear way \cite{Garla2004}.
So, by controlling for both strengths and degrees, our approach is indirectly controlling for the GDP of countries as well.
The surprising agreement between our model and the data does not however imply that the other aforementioned dyadic factors (involving pairs of countries) are unimportant. 
Rather, our analysis shows that, among the country-specific factors, the ones that only affect the strengths are definitely less informative than those that impact both strengths and degrees.
Adding the distances, or other dyadic factors, can in principle lead to an even better agreement between model and data.
On the other hand, it is interesting to notice that geographic distances are sometimes outperformed by purely topological properties (such as the reciprocity \cite{picciolo2012role}) in explaining the structure of the WTW, and in general Gravity Models are much less effective than maximum-entropy ensembles in reproducing the binary structure of the WTW \cite{duenas2011modeling,myTinbergen}. 
The main reason for this ineffectiveness is that, depending on their specification, Gravity Models tend to predict a network that is either too dense (in the simplest setting, even fully connected \cite{myTinbergen}) or too topologically different from the observed WTW \cite{duenas2011modeling}.  

\subsection{Intensive and extensive margins}
Another series of important economic considerations concerns the relationship between our findings and the so-called \emph{extensive} and \emph{intensive margins} of trade. 
These two concepts, first introduced by Ricardo \cite{ricardo1819principles}, are widely used in economics, and should not be confused with the notion of intensive and extensive variables in statistical physics and thermodynamics. 
In the context of international trade,
extensive and intensive margins refer to tendency of the network to evolve through the creation of new trade connections or through the reinforcement of existing ones  respectively \cite{felbermayr2006exploring,DeBene_Tajoli_2011}.

Even if both margins are known to be relevant, neither a systematic treatment of their role in the prediction of international trade relationships, nor an agreement on their relative importance can be found in the economic literature.
Some works agree on the relevance of extensive margins. For instance, a cross-country analysis reveals that extensive margins account for the $60\%$ of exports of the larger economies \cite{hummels2002variety}, and another study shows that increasing extensive margins means augmenting exports of developing countries \cite{evenett2009export}.

At the same time, a  large body of work stresses the relevance of intensive margins. 
For instance, intensive margins represented a fundamental factor in the period 1970-1990 \cite{felbermayr2006exploring,helpman2008estimating} and have been particularly relevant for China's exports growth in the period 1992-2005 \cite{amiti2010anatomy} and for Colombian countries' exports \cite{eaton2008margins}. 
It has also been claimed \cite{besedevs2011role} that the majority of trade growth is due to the intensive margin, rather than the extensive one. 
This view stresses the importance to focus on a dynamical comparison (e.g. introducing the concepts of `survival' and `deepening' to characterize export relationships) rather than on a standard static approach.

The controversial results existing in the literature appear to be mainly due to the different levels at which the two margins are examined \cite{besedevs2011role}: some works define extensive margins at the country-product level, others at the product level, others at the country level. 
This problem, together with the composite effect of international changes on trade, can determine mixed and contradictory results. For example, trade liberalization affects trade flows in two ways. On the one hand, since trade becomes less costly, the trade volumes increase (intensive margin at product level). 
On the other hand, more firms trade and more goods are traded (extensive margin both at product and country-product level). 
It has been observed \cite{chaney2008distorted} that the elasticity of substitution should also be taken into account, because it has opposite effects on the two margins: high elasticity makes the intensive margin more sensitive to changes in trade barriers (trade costs), whereas the extensive margin is less sensitive to this effect. 

To the best of our knowledge, in the economic literature there has been no systematic analysis of the predictive power of extensive and intensive trade margins so far. 
Starting from one snapshot of the international trade network, is the knowledge of the growth of trade along the intensive and/or extensive margin enough to predict the structure of the network at a later time?

Even if our results cannot fully answer such question, they suggest a plausible scenario. 
We first  note that a change in the degree (number of partners) of a country implies that the network is evolving along the extensive margin of trade. 
On the other hand, a change in the strength (total volume) can be either be due to changes in the number of partners or to changes of the amount of trade for existing links. 
This means that, while changes in the degree only reflect the extensive margin, changes in the strength reflect both the extensive and intensive margins: this is a second asymmetry between the different pieces of information encoded into the degrees and the strengths.
It is also another indication that the WCM, by enforcing the strengths alone, cannot distinguish between the two margins, while the ECM can isolate the extensive information (degrees) from the combined one (strengths).

Our findings imply that, if the structure of the international trade network is known at time $t$, and if the growth (or decrease) of both strengths and degrees from time $t$ to time $t+\Delta t$ is also known, then it is possible to predict the structure of the network at time $t+\Delta t$ with great accuracy. 
By contrast, if only the growth of the strengths is known, the future structure of the network cannot be satisfactorily predicted. 
These results can be interpreted in terms of the fact that a combined knowledge of intensive and extensive margins (in this case, the change of the strengths) does not allow us to correctly model the network, while if the extensive margin (change of the degrees) is also separately specified (thus indirectly controlling for the residual intensive margin as well), then the model can successfully explain the data.

\subsection{Intensive and extensive biases}
We finally show that our results offer additional and important interpretations about the intensive/extensive dichotomy, in a way that differs from the traditional one in macroeconomics.
While the economic literature has mostly tried to quantify the extent to which international trade has evolved along each of the two margins, with the purpose of identifying the most important direction of trade growth, our results naturally suggest a novel, intrinsically static perspective.

It should be noted that the ECM specified by eq.(\ref{qw}) has an important abstract property. It is mathematically equivalent to a network formation process where, between any two (initially disconnected) nodes $i$ and $j$, a link of unit weight is first established with probability $p_{ij}$ given by eq.(\ref{pij}), and then (if the previous attempt is successful) strengthened by another unit of weight with probability $y^*_iy^*_j$.
For each pair of vertices, such weight-increasing attempts are iterated with the same probability $y^*_iy^*_j$ if the previous attempt was successful, and stop as soon as a failure occurs. 
This means that the probability of establishing a unit link for the first time is $p_{ij}$, while that of reinforcing an existing link by another weight unit is $y^*_i y^*_j$.

Now, it is easy to show that $p_{ij}> y^*_i y^*_j$ if and only if $x^*_i x^*_j>1$.
Therefore, if $x^*_i x^*_j > 1$ ($x^*_i x^*_j < 1$) the creation a link of unit weight between nodes $i$ and $j$ has a larger (smaller) probability than the reinforcement of the same link by a unit of weight.
This feature makes the model particularly appropriate to study the extensive/intensive dichotomy in a novel sense, as the value of $x^*_i x^*_j$ can bias the network, at a purely static level, towards the extensive ($x^*_i x^*_j>1$) or the intensive ($x^*_i x^*_j<1$) direction. 

More in general, in the network formation process the probability of establishing a link of weight $w$ between two previously disconnected vertices (irrespective of possibile further reinforcements) is $p_{ij} (y^*_i y^*_j)^{w-1}$, while that of adding a weight $w$ (again, irrespective of possible further increases) to an already existing connection is $(y^*_i y^*_j)^w$. 
In this case as well, the former probability is larger than the latter if and only if $x^*_i x^*_j>1$.
So, independently of the value of $w$, $x^*_i x^*_j>1$ implies a tendency towards the extensive direction, while  $x^*_i x^*_j<1$ signals a preference for the intensive one.

For the above reasons, we denote $x^*_i x^*_j$ as the `extensive bias' for the pair $i,j$.
Note that, since the extensive bias is a product of two country-specific values, it is not possible to determine, on the basis of the value of $x_i^*$ for a single country, whether the dominant bias for that country is the extensive or the intensive one. Thus the preference for one bias turns out to be an inherently dyadic property.

If $x^*_i x^*_j=1$ for all $i,j$, then the network is neutral with respect to link creation and link reinforcement. Now, it should be noted that this is precisely what is obtained in the WCM (where only the strengths are specified) because, as we mentioned, the latter can be regarded as a particular case of the ECM where $x_i=1$ for all $i$ \cite{mastrandrea2013enhanced}.
In other words, the WCM assumes that the network is neutral with respect to the two biases, and that there is no preference between the extensive and intensive direction.
By contrast, the ECM assumes that, for each pair of nodes, there can be a different kind of bias.
Since we found that the WCM and the ECM perform very bad and very good respectively, we have a strong empirical indication that the WTW is \emph{not} bias-neutral.

The extensive bias $x_i^* x_j^*$ indicates the preference of a specific pair of countries for the dominant direction, as measured on a particular snapshot/layer of the WTW.
This notion of extensive or intensive bias should therefore not be interpreted in the same sense as the extensive or intensive margin, i.e. as a preferred direction for the dynamical evolution of the network, but in terms of the `static' deviation of the real network (well reproduced by the ECM) from the neutral topology expected under the WCM.
In this sense, the WCM is serving as a null model indicating how an economic network would look like if the extensive and intensive biases were balanced.

These considerations lead us to interpret that, in order to reproduce the observed structure of the WTW, we need to enforce realistic extensive and intensive biases as detected by the ECM through the additional knowledge of the degrees. From the strengths alone, it is indeed impossible to infer the bias towards a specific direction.

\section{Concluding Remarks}

In this paper we employed a maximum-entropy approach to the WTW, which is an important example of economic multiplex. 
From a theoretical point of view, our findings completely reverse the standard results concerning the reconstruction of weighted networks from local node-specific information. 
We proved that it is indeed possible to reproduce at a highly satisfactory level several higher-order binary and weighted properties for the WTW, provided that the enforced local constraints include both strenghts and degrees.
Our results are robust across different levels of disaggregation and several temporal snapshots.

Economically speaking, these and previous results \cite{Squartini_etal_2011a_pre,Squartini_etal_2011b_pre,fagiolo2013null} allow us to make some considerations in relation to the extensive and intensive margins of trade, in a novel `static' fashion.
We have shown that the specification of the strengths, without the separate specification of the degrees, corresponds to the assumption that (at a static level) the system is neutral with respect to the two margins, i.e. the extensive and intensive biases are perfectly balanced.
By contrast, if the degrees are also specified, then for every pair of vertices there is a specific tendency to favour one of the two directions.
The fact that the latter model reproduces the real WTW very well, while the former performs very bad, is then a clear indication that the network is not neutral with respect to the two biases.
Without specifying the extensive biases the graph would be almost fully connected, while without specifying the intensive ones we would not be able to predict the magnitude of connections. This is the reason why both intensive and extensive biases are needed.
We also found that different pairs of countries have different intrinsic biases towards either the extensive of the intensive direction. 
If such dyadic biases are not taken into account, explaining the observed structure of the WTW appears impossible.

Despite its static character, our analysis allows us to draw some interesting implications on the predictive power of trade margins also from a dynamic perspective. 
However, a more rigorous verification of the relationship between trade margins and extensive/intensive biases through the exploration of a complementary, explicitly dynamic framework is an important step to take in future research. 

Since the effectiveness of the ECM has been shown for various other networks including non-economic ones \cite{mastrandrea2013enhanced}, the importance of separately specifying the extensive and intensive biases might actually be a very general result.
Moreover, our findings represent the first (to the best of our knowledge) empirical evidence in favour of the idea that, layer by layer, real-world multiplexes can be strongly shaped by local constraints.

\acknowledgements
This work was supported by the EU project MULTIPLEX (contract 317532) and the Netherlands Organization for Scientific Research (NWO/OCW).
D.G. acknowledges support from the Dutch Econophysics Foundation (Stichting Econophysics, Leiden, the Netherlands) with funds from beneficiaries of Duyfken Trading Knowledge BV, Amsterdam, the Netherlands.

\bibliography{resubmitted2PRE.bbl}

\end{document}